\documentclass[twocolumn]{article}
\let\ifapj\iffalse
\let\ifarxiv\iftrue
\let\iflocal\iffalse
\ifapj\usepackage{patch-apj}\NewPageAfterKeywords\fi
\usepackage{etoolbox,ifxetex,ifluatex}
\ifxetex\else\ifluatex\else\usepackage[utf8]{inputenc}\fi\fi
\ifluatex\AtEndPreamble{\hypersetup{pdfencoding=auto}}\fi
\usepackage{csquotes,siunitx}
\ifapj
  \usepackage{acro,appendix,hyperref}
  \acsetup{patch/longtable=false}
  
  \let\symsf\mathsf
  \let\symcal\mathcal
\else\ifarxiv
  \usepackage{tikz}
\else\iflocal
  \usepackage{tikz}
  \usetikzlibrary{external}
\fi\fi\fi
\makeatletter\ifboolexpr{bool{arxiv} or bool{local}}{
  \usepackage{font-termes,draft}
  \usepackage[physics, astro]{basic}
  \usepackage[mode=biblatex, colorlinks]{apjbib}
  \ExecuteBibliographyOptions{embedlinks}
  \addbibresource{mri.bib}
  \renewcommand*\incr@eqnum
    {\refstepcounter{equation}\ifmmode\mathopen\fi{}\let\incr@eqnum\@empty}
  \hypersetup{urlcolor=blue}
}{}\makeatother
\DeclareAcronym{MRI}{
  short           =MRI,
  long            =magnetorotational instability,
  short-indefinite=an}
\DeclareAcronym{GR}{
  short           =GR,
  long            =general-relativistic}
\DeclareAcronym{MHD}{
  short           =MHD,
  long            =magnetohydrodynamic,
  short-indefinite=an,
  short-plural    =,
  long-plural     =s}
\DeclareAcronym{GRMHD}{
  short           =GRMHD,
  long            =general-relativistic magnetohydrodynamic,
  short-plural    =,
  long-plural     =s}
\DeclareAcronym{TDE}{
  short           =TDE,
  long            =tidal disruption event}
\DeclareAcronym{UV}{
  short           =UV,
  long            =ultraviolet,
  long-indefinite =an}
\newcommand*\astF{\mathord\ast F}
\usepackage{xcolor}
\expandafter\def\csname editcolor1\endcsname{blue!60!cyan}
\ifapj\let\edit\relax\fi
\newcommand*\edit[2]{\textcolor{\csname editcolor#1\endcsname}{#2}}
\begin{document}

\title{Three-dimensional simulations of the
\texorpdfstring\\\ magnetorotational instability in eccentric disks}

\ifapj
  \author[0000-0001-5949-6109]{Chi-Ho Chan}
  \affiliation{Center for Relativistic Astrophysics and School of Physics,
  Georgia Institute of Technology, Atlanta, GA 30332, USA}
  \author[0000-0002-7964-5420]{Tsvi Piran}
  \affiliation{Racah Institute of Physics, Hebrew University of Jerusalem,
  Jerusalem 91904, Israel}
  \author[0000-0002-2995-7717]{Julian~H. Krolik}
  \affiliation{Department of Physics and Astronomy, Johns Hopkins University,
  Baltimore, MD 21218, USA}
\fi

\ifboolexpr{bool{arxiv} or bool{local}}{
  \author{Chi-Ho Chan}
  \affil{Center for Relativistic Astrophysics and School of Physics,
  Georgia Institute of Technology, Atlanta, GA 30332, USA}
  \author{Tsvi Piran}
  \affil{Racah Institute of Physics, Hebrew University of Jerusalem,
  Jerusalem 91904, Israel}
  \author{Julian~H. Krolik}
  \affil{Department of Physics and Astronomy, Johns Hopkins University,
  Baltimore, MD 21218, USA}
}{}

\date{June 25, 2024}
\ifapj\else\uats{%
  Magnetohydrodynamical simulations (1966);
  Accretion (14);
  Black hole physics (159);
  Gravitation (661)}
\fi

\shorttitle{3D eccentric MRI}
\shortauthors{Chan et al.}
\pdftitle{Three-dimensional simulations of the magnetorotational instability in
eccentric disks}
\pdfauthors{Chi-Ho Chan, Tsvi Piran, Julian H. Krolik}

\begin{abstract}
Previously we demonstrated that the \ac{MRI} grows vigorously in eccentric
disks, much as it does in circular disks, and we investigated the nonlinear
development of the eccentric \ac{MRI} without vertical gravity. Here we explore
how vertical gravity influences the \ac{MHD} turbulence stirred by the
eccentric \ac{MRI}. Similar to eccentric disks without vertical gravity, the
ratio of Maxwell stress to pressure, or the Shakura--Sunyaev $\alpha$
parameter, remains \num{\sim e-2}, and the local sign flip in the Maxwell
stress persists. Vertical gravity also introduces two new effects. Strong
vertical compression near pericenter amplifies reconnection and dissipation,
weakening the magnetic field. Angular momentum transport by \ac{MHD} stresses
broadens the mass distribution over eccentricity at much faster rates than
without vertical gravity; as a result, spatial distributions of mass and
eccentricity can be substantially modified in just \numrange{\sim5}{10} orbits.
\Ac{MHD} stresses in the eccentric debris of \aclp{TDE} may power emission
\qty{\gtrsim1}{\year} after disruption.
\end{abstract}
\acresetall

\section{Introduction}

Much has been learned about the dynamics of disks whose material travels on
circular orbits. In disks with enough ionization to support electrical
conductivity, the internal stress transporting angular momentum outward and
allowing material to drift inward is primarily magnetic: The fast-growing
\ac{MRI} stirs up \ac{MHD} turbulence, and orbital shear correlates the radial
and azimuthal components of the turbulent magnetic field to produce this stress
\citep{1991ApJ...376..214B, 1998RvMP...70....1B, 1991ApJ...376..223H}. The
nonlinear development of the \ac{MRI} has been extensively explored, both in
shearing-box and global-disk settings \citep[e.g., most
recently,][]{2022A&A...659A..91W, 2022MNRAS.517.2639Z, 2024MNRAS.530.1866S,
2024MNRAS.532.1522J}.

However, disks might also be eccentric. One common cause is external
gravitational perturbation: Disks acquire forced and free eccentricities as a
result of secular gravitational interaction with eccentric binaries
\citep[e.g.,][]{2000ssd..book.....M}, and nonaxisymmetric components of gravity
resonantly amplify small initial eccentricities \citep{1988MNRAS.232...35W,
1991ApJ...381..259L}. Internal processes such as viscous overstability
\citep{1978MNRAS.185..629K, 1994MNRAS.266..583L, 2001MNRAS.325..231O} can also
give rise to eccentricity. Alternatively, disks can be created eccentric in
multiple ways: outgassing from planetesimals \citep{2021MNRAS.505L..21T}, or
tidal disruption of stars \citep{2015ApJ...804...85S, 2015ApJ...806..164P,
2017MNRAS.467.1426S} and molecular clouds \citep[e.g.,][]{2008Sci...321.1060B}
by supermassive black holes. There is observational evidence for eccentric
disks as well, such as asymmetric lines in white dwarfs
\citep[e.g.,][]{2006Sci...314.1908G}, and asymmetric broad emission lines in
active galactic nuclei \citep[e.g.,][]{1995ApJ...438..610E,
2021MNRAS.506.6014T} and \acp{TDE} \citep[e.g.,][]{2014ApJ...783...23G,
2017MNRAS.472L..99L}.

In magnetized eccentric disks, \ac{MHD} stresses may outcompete hydrodynamic
forces in directing disk evolution. Analytic treatment of stresses in eccentric
disks \citep{2001MNRAS.325..231O, 2021MNRAS.500.4110L, 2021MNRAS.501.5500L} is
limited in applicability because the prescriptions it relies on, even when
applied to circular disks, may not adequately represent the nature of turbulent
\ac{MHD} stresses pumped by the \ac{MRI}. Recent years have seen exploration of
the \ac{MRI} in eccentric disks. The studies published so far all assume
unstratified disks, that is, disks unaffected by vertical gravity.
\Citet{2018ApJ...856...12C} showed analytically that the \ac{MRI} is a robust
instability that continues to be active up to extremely high eccentricities.
\Citet{2022ApJ...933...81C} conducted \acp{MHD} simulations to examine how the
variation of orbital properties around the disk affects the nonlinear evolution
of the \ac{MRI} in moderately eccentric disks. These simulations revealed that
the \ac{MRI} develops much like in circular disks, but with noticeable local
departures. Interestingly, the simulations of \citet{2020MNRAS.497..451D}
suggested that sharp changes in eccentricity or orientation over radius may
suppress the \ac{MRI}.

However, the absence of vertical gravity in unstratified disks is unphysical.
Vertical gravity changes how the gas is distributed, modifying the pressure
distribution even in hydrodynamic disks and altering how \acp{MHD} operates in
magnetized disks. For these reasons, this article investigates the behavior of
the eccentric \ac{MRI} in disks subject to vertical gravity, also known as
stratified disks. Vertical gravity affects eccentric disks in ways not possible
in circular disks. As material in an eccentric disk moves closer and farther
away from the central object, vertical gravity strengthens and weakens, and the
disk collapses and puffs up accordingly. Vertical gravity varies on the orbital
timescale, which is comparable to the vertical sound-crossing timescale, so the
disk is never in vertical force balance. The amplitude of this vertical disk
breathing can be quite large even at moderate eccentricities
\citetext{\citealp{2014MNRAS.445.2621O, 2021ApJ...920..130R}\multicitedelim but
see \citealp{2021MNRAS.501.5500L}}. Three-dimensional effects can manifest
themselves hydrodynamically \citep{2008MNRAS.388.1372O, 2016MNRAS.458.3221T,
2021ApJ...920..130R}; as we shall show here, \ac{MHD} effects can be
considerably stronger.

Our previous unstratified simulations \citep{2022ApJ...933...81C} were novel in
that, although their dynamics were purely Newtonian, they took advantage of the
coordinate flexibility afforded by \iac{GR} formulation to implement a
coordinate system tailored to the eccentric disk shape. Here we build on that
framework: The strong height modulation makes a stratified disk challenging to
resolve with traditional grids, so we generalize the method to devise a
customized grid with variable cell height.

We explain our calculational approach in \cref{sec:methods}; a high-level
synopsis is given in \cref{sec:overview}. Our results are collected in
\cref{sec:results}, prefaced by a summary of key findings in
\cref{sec:summary}. We discuss these results in \cref{sec:discussion} and
conclude in \cref{sec:conclusions}.

\section{Methods}
\label{sec:methods}

\subsection{Overview}
\label{sec:overview}

We study the nonlinear development of the \ac{MRI} in a stratified disk with a
moderate eccentricity that is uniform across the disk. This work extends the
unstratified simulations in \citet{2022ApJ...933...81C} by adding vertical
gravity. The disk height responds to the varying vertical gravity along the
orbit; our goal is to see how this disk breathing influences \ac{MHD}
turbulence. For a uniformly eccentric disk, midplane orbits are divergence-free
and surface density is constant along any orbit; the lack of horizontal
compression makes this kind of disk useful for isolating the effects of
vertical gravity.

It is instructive to situate our disk in the context of previous studies of
eccentric disks. Two fundamental characteristics of an eccentric disk are its
eccentricity and aspect ratio, the latter being the ratio of vertical scale
height to orbital length scale. Our disk is moderately eccentric and
geometrically thin, resembling those studied analytically by, e.g.,
\citet{2014MNRAS.445.2621O}. \Citet{2021ApJ...920..130R} found that vertical
gravity promotes shocks in the case of high eccentricity and large geometrical
thickness, but neither criterion for shocks is satisfied in our disk.

We assume the gas is adiabatic with an adiabatic index of $\gamma=\tfrac53$, in
contrast to the almost isothermal equation of state used in
\citet{2022ApJ...933...81C}. A stiffer equation of state reduces the degree of
disk breathing, making the problem more numerically tractable, but has little
effect on the character of \ac{MHD} turbulence. We also assume that ideal
\acp{MHD} applies.

Despite the Newtonian nature of the problem, we follow
\citet{2022ApJ...933...81C} and adopt a \acp{GRMHD} formulation so we can use
the coordinate freedom that comes with it to make our grid better match the
geometry of a stratified eccentric disk. One reason we employ the finite-volume
\acp{GRMHD} code Athena++ \citep{2016ApJS..225...22W, 2020ApJS..249....4S} is
that it readily supports arbitrary coordinate systems. In keeping with \ac{GR}
tradition, we adopt gravitational units, that is, the central mass $M$ is
unity, velocities are in terms of the speed of light $c$, and lengths are in
gravitational radii $GM/c^2$.

As \citet{2022ApJ...933...81C} argued, the balance between the competing
desires to have Newtonian dynamics and to limit truncation errors leads us to
consider a disk with a characteristic semimajor axis of $a_*=200$. To allow the
\ac{MRI} ample time to grow, the disk should not evolve too rapidly under
hydrodynamics; therefore, we consider a moderate eccentricity of $e=0.4$, a bit
smaller than the $e=0.5$ used in \citet{2022ApJ...933...81C}. These choices
imply a characteristic semilatus rectum $\lambda_*=a_*(1-e^2)=168$. In the same
vein, we assign to our initial disk a small sound speed, equivalent to a Mach
number of \num{\approx30} at pericenter.

\subsection{Equations}

We follow the conventions of \citet{2003ApJ...589..444G}. The metric signature
is $(-,+,+,+)$, Greek indices range over $\{0,1,2,3\}$, Latin indices range
over $\{1,2,3\}$, and Einstein summation is implied. The equations of
\acp{GRMHD} are
\begin{alignat}{3}
& \partial_t[(-g)^{1/2}\rho u^t
  &]& +\partial_j[(-g)^{1/2}\rho u^j
  &]&= 0, \\
& \partial_t[(-g)^{1/2}T^t_\mu
  &]& +\partial_j[(-g)^{1/2}T^j_\mu
  &]&= (-g)^{1/2}T^\nu_\sigma\Gamma^\sigma_{\mu\nu}, \\
& \partial_t[(-g)^{1/2}\astF^{it}
  &]& +\partial_j[(-g)^{1/2}\astF^{ij}
  &]&= 0.
\end{alignat}
Here $t$ is the coordinate time, $\rho$ is the comoving mass density, $u^\mu$
is the velocity, $g$ is the determinant of the metric $g_{\mu\nu}$, and
$\Gamma^\sigma_{\mu\nu}$ is the Christoffel symbol of the second kind. From the
Hodge dual $\astF^{\mu\nu}$ of the electromagnetic tensor $F^{\mu\nu}$ we
obtain the magnetic field $B^i=\astF^{i0}$ and the projected magnetic field
$b^\mu=u_\nu\astF^{\nu\mu}$. Lastly, the stress--energy tensor is
\begin{equation}
T^{\mu\nu}=\biggl(p+\frac12b_\sigma b^\sigma\biggr)g^{\mu\nu}
  +\biggl(\rho+\frac\gamma{\gamma-1}p+b_\sigma b^\sigma\biggr)u^\mu u^\nu
  -b^\mu b^\nu,
\end{equation}
where $p$ is the gas pressure. Our analysis makes use of the current, which is
not tracked explicitly by the code but can be obtained in post-processing from
the expression
\begin{equation}
j^\mu=\nabla_\nu F^{\mu\nu}
  =\nabla_\nu(\epsilon^{\mu\nu\sigma\tau}u_\sigma b_\tau),
\end{equation}
with $\epsilon^{\mu\nu\sigma\tau}$ the contravariant Levi--Civita tensor.

\subsection{Eccentric coordinate system}
\label{sec:coordinate system}

We employ a three-dimensional coordinate system based on conventional
cylindrical coordinates, modified to conform to the gas motion in an eccentric
disk. The coordinate curves in the midplane are particle orbits with the
initial disk eccentricity, and the expansion and contraction of the vertical
coordinate match that of the initial disk. Using such a coordinate system has
multiple benefits. It is more efficient than the standard cylindrical
coordinate system because the simulation domain fits more tightly around the
disk, and because there is no need to allocate cells to regions far away from
the disk. Cells are shorter near pericenter where the disk is thinner and
taller near apocenter where the disk is thicker, so the level to which disk
mass is resolved is more azimuthally uniform. Streamlines are exactly parallel
to grid lines in the initial disk and remain roughly so over most of the orbit
even after the disk has evolved differentially in eccentricity and orientation;
the lower obliqueness of the streamlines relative to the grid reduces numerical
dissipation and numerical artifacts, particularly during the early-time linear
stage of the \ac{MRI}. The initial velocity and magnetic field can also be set
in a much cleaner fashion, as we shall see in \cref{sec:initial condition}.

The eccentric coordinate system is derived from the cylindrical coordinate
system in two orthogonal steps: first, by substituting cylindrical coordinate
surfaces with elliptical ones that match the initial disk eccentricity; second,
by deforming horizontal coordinate surfaces into oblique conical ones that
approximately track disk breathing. The intersection of the elliptical
coordinate surfaces with the midplane are orbits followed by material in the
initial disk.

\begin{figure}
\ifapj\includegraphics{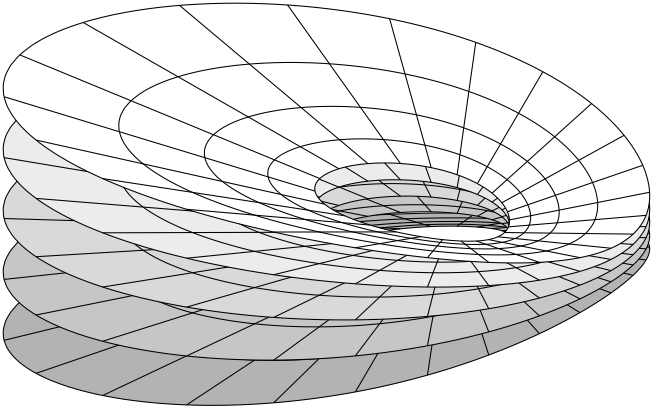}\else
\usetikzlibrary{perspective}
\begin{tikzpicture}[3d view={20}{20}, scale=2]
\pgfmathsetmacro\e{0.4}
\pgfmathsetmacro\ld{0.3}
\pgfmathsetmacro\ll{-2}
\pgfmathsetmacro\lu{2}
\pgfmathsetmacro\pd{15}
\pgfmathsetmacro\pl{-180/\pd}
\pgfmathsetmacro\pu{180/\pd-1}
\pgfmathsetmacro\zd{0.2}
\pgfmathsetmacro\zl{-2}
\pgfmathsetmacro\zu{2}
\newcommand*\elr[3]{((#2)/(1+(#1)*cos(#3)))}
\newcommand*\elx[3]{(\elr{#1}{#2}{#3}*cos(#3))}
\newcommand*\ely[3]{(\elr{#1}{#2}{#3}*sin(#3))}
\newcommand*\grid{
  \begin{scope}
  \foreach \z in {\zl,...,\zu} {
    \pgfmathsetmacro\f{(\zu-\z)/(\zu-\zl)*30}
    \pgfmathsetmacro\z{\z*\zd}
    {\pgfmathsetmacro\ll{exp(\ll*\ld)}
     \pgfmathsetmacro\lu{exp(\lu*\ld)}
     \fill[variable=\p, domain=-180:180, samples=100, smooth,
       black!\f, even odd rule]
       plot ({\elx\e\ll\p},{\ely\e\ll\p},{\sh\ll\p\z})
       plot ({\elx\e\lu\p},{\ely\e\lu\p},{\sh\lu\p\z});}
    \foreach \l in {\ll,...,\lu} {
      \pgfmathsetmacro\l{exp(\l*\ld)}
      \draw[variable=\p, domain=-180:180, samples=100, smooth]
        plot ({\elx\e\l\p},{\ely\e\l\p},{\sh\l\p\z});}
    \foreach \p in {\pl,...,\pu} {
      \pgfmathsetmacro\p{\p*\pd}
      \pgfmathsetmacro\ll{exp(\ll*\ld)}
      \pgfmathsetmacro\lu{exp(\lu*\ld)}
      \draw
        ({\elx\e\ll\p},{\ely\e\ll\p},{\sh\ll\p\z}) --
        ({\elx\e\lu\p},{\ely\e\lu\p},{\sh\lu\p\z});}}
  \end{scope}}
\begin{scope}
\newcommand*\sh[3]{(0.2+0.5*(1-cos(#2)))*(#1)*(#3)}
\grid
\end{scope}
\end{tikzpicture}\fi
\caption{Coordinate surfaces of constant $\zeta$ for one instance of the
eccentric coordinate system.}
\label{fig:grid}
\end{figure}

The eccentric coordinate system is parametrized by $(e,\lambda_*)$, where $e$
is the initial disk eccentricity and $\lambda_*$ is the characteristic
semilatus rectum of the orbit along which we compute the expected amplitude of
disk breathing for use in constructing the grid. One instantiation of the
eccentric coordinate system is illustrated in \cref{fig:grid}.

More explicitly, let $(t,R,\varphi,z)$ be cylindrical coordinates. We assume
the weak-gravity limit, so the only nonzero components of the metric are
\begin{align}
g_{tt} &= -(1+2\Phi), \\
g_{RR} &= 1, \\
g_{\varphi\varphi} &= R^2, \\
g_{zz} &= 1.
\end{align}
\Citet{2022ApJ...933...81C} proved that closed elliptical orbits exist at all
semilatera recta for the quasi-Newtonian potential
\begin{equation}
\Phi(R,z)=-1/[(R^2+z^2)^{1/2}+2].
\end{equation}
Eccentric coordinates $(t,\ln\lambda,\phi,\zeta)$ are derived from cylindrical
coordinates through the coordinate transformation
\begin{align}
\label{eq:transform 1}
R &= \lambda/(1+e\cos\phi), \\
\label{eq:transform 2}
\varphi &= \phi, \\
\label{eq:transform 3}
z &= \zeta H_*(e;\phi)\lambda/\lambda_*.
\end{align}
We can reuse $t$ for the new coordinates without ambiguity because it does not
enter into the coordinate transformation.

Logarithmic scaling in the semilateral rectus direction is built into the
eccentric coordinate system. The rest of this article will not mention the
($\ln\lambda$)\nobreakdash-coordinate itself; measurements along that direction
will be converted to $\lambda$ first for ease of comprehension. A caveat is
that whenever a component is labeled with $\lambda$, it is understood that the
($\ln\lambda$)\nobreakdash-coordinate is referred to. Grid breathing is
controlled by the arbitrary grid height function $H_*$, which should reduce to
a constant function when $e=0$. The metric and connection are listed in
\cref{sec:metric}.

One way of fixing $H_*$ utilizes a reference gas column that mimics the initial
disk. The column obeys the same adiabatic equation of state
$p\propto\rho^\gamma$ as the disk, is isentropic for simplicity, and has
surface density $\Sigma_*$ and entropy typical of the initial disk at the
characteristic orbit $\lambda=\lambda_*$. As we shall see in
\cref{sec:deformation}, the surface density of our initial disk depends only on
$\lambda$, so the choice of $\Sigma_*$ is straightforward. However, because our
initial disk is not in general isentropic, the proportionality constant $K_*$
in the equation of state can be only loosely related to the initial disk
entropy, and \cref{sec:deformation} will explain how it is picked. The column
so constructed is placed at different $\phi$ along the characteristic orbit,
and the scale height of the column under vertical force balance is our
$H_*(\phi)$. The details are relegated to \cref{sec:grid height}. Because the
location of the column is determined by $(e,\lambda_*,\phi)$ and the gas in the
column is parametrized by $(\gamma,\Sigma_*,K_*)$, the full dependence of $H_*$
is $H_*(e,\lambda_*,\gamma,\Sigma_*,K_*;\phi)$.

We emphasize that the choice of $H_*$ is arbitrary and has no physical
implications. The breathing of our grid captures only part of the breathing of
the actual disk: Our $H_*$ is based on an adiabatic hydrostatic column, whereas
vertical motion in an actual disk exhibits inertia, and the accumulation of
dissipated energy can also puff up the disk. We could use more elaborate
methods to determine $H_*$ \citep[e.g.,][]{2014MNRAS.445.2621O}, but our
approximate prescription has the advantage that $H_*$ and its derivatives have
simple expressions.

\subsection{Initial condition}
\label{sec:initial condition}

\begin{figure}
\includegraphics{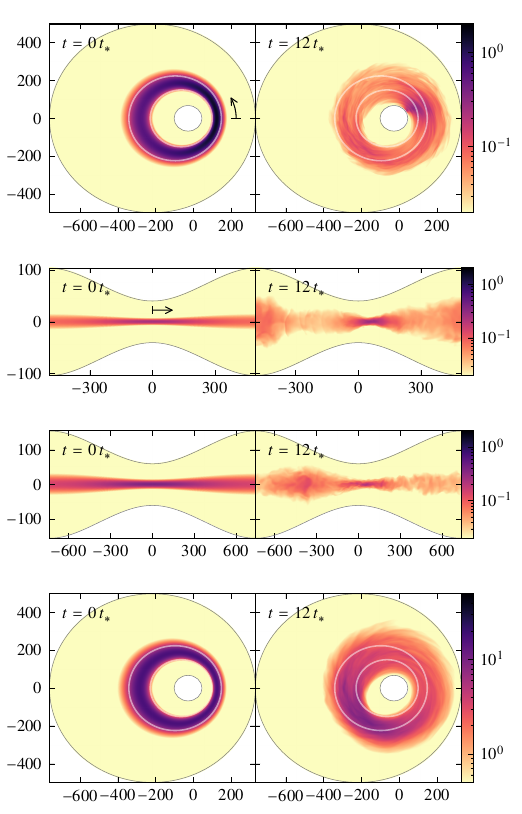}
\caption{Density in the top three rows and surface density in the bottom row.
The top row shows midplane slices of density, the middle two rows show vertical
slices of density along $\ln(\lambda/\lambda_*)=\mp0.2$, respectively, on
either side of the characteristic orbit, and the bottom row shows surface
density. The boundaries of the simulation domain are traced by thin gray
curves. For the top and bottom rows, the ellipses along which vertical slices
are taken are indicated by faint white ellipses. For the middle two rows, the
abscissa is the physical arc length along the orbit measured from $\phi=0$, the
ordinate is the physical height $z$ above the midplane, and the ordinate is
more stretched than the abscissa. The two arrows in the top two rows indicate
how $\phi$ is defined: The tick is where $\phi=0$ and the arrow points in the
direction of increasing $\phi$.}
\label{fig:density}
\end{figure}

We would like to start our simulation with an eccentric disk as close to force
balance as possible, both horizontally and vertically. To do so, we deform a
circular disk that is strictly in hydrostatic equilibrium into an eccentric
disk that is approximately so. The result is exhibited in the left column of
\cref{fig:density}. What follows is a sketch of the deformation procedure;
technicalities are left to \cref{sec:deformation}.

Our starting point is a circular disk. The density maximum of the disk is along
$R=\lambda_*$, where $\lambda_*=168$ is the characteristic semilatus rectum of
the eccentric disk we would like to arrive at. We match the circular and
eccentric disks in semilatus rectum in order that the two disks have the same
characteristic specific angular momentum. The circular disk is both
horizontally and vertically thin: Starting from the density maximum, the
density falls by one $e$\nobreakdash-folding if we move to
$R\approx0.86\,\lambda_*$ or $R\approx1.18\,\lambda_*$ along the midplane, or
to $\abs z\approx9$ vertically. The Mach number of the disk is \num{\approx30}.

The circular disk is converted into an initial condition. The initial condition
uses a realization of the eccentric coordinate system adapted to the disk, with
$e=0$ and a constant $H_*(e=0;\phi)$ matched to disk conditions. We do not use
the initial condition directly; instead, with the circular disk already set up
in coordinate space, we turn it eccentric simply by changing $e$ of the
coordinate system to some finite value. This reinterpretation of the initial
condition simultaneously makes $H_*(e;\phi)$ a function of $\phi$. Because the
density distribution of the disk follows coordinate surfaces, the poloidal
cross-section of the disk expands and contracts over azimuth the same way the
grid does. Furthermore, the varying cross-section implies the velocity should
have a nonzero vertical component, which our procedure guarantees because the
velocity stays parallel to coordinate curves.

We paint on the eccentric disk a dipolar magnetic field derived from a magnetic
potential whose only nonzero component is
\begin{equation}
A_\phi(\lambda,\phi,\zeta)\propto
  (-g)^{1/2}
  \max\{0,\rho-\tfrac12\max_{\lambda',\zeta'}\rho(\lambda',\phi,\zeta')\};
\end{equation}
for intuition, $\uvec e^\phi=\uvec e^\varphi$ is a one-form that can be
visualized as a family of poloidal planes. Including the metric determinant in
the magnetic potential makes the magnetic field strength more uniform over
azimuth, and using the maximum density in the same poloidal slice instead of
the global maximum ensures all poloidal planes are magnetized. The
proportionality constant is picked such that the volume-integrated plasma beta,
or the ratio of volume-integrated gas pressure to volume-integrated magnetic
pressure, is 100. After the magnetic field is added, we reduce the gas pressure
so as to keep the total pressure the same as before, and we perturb the gas
pressure at the 0.01 level to seed the \ac{MRI}.

\subsection{Simulation domain, boundary conditions, and other numerical
concerns}

The simulation domain occupies the volume
$[\exp(-1)\lambda_*,\allowbreak\exp(1)\lambda_*]\times\allowbreak
[-\pi,\pi]\times\allowbreak[-6,6]$ in $(\lambda,\phi,\zeta)$. The vertical
extent at the characteristic orbit needs to be six times the scale height of
the reference column to capture the gas and magnetic field at high altitudes.
The grid is linear in coordinate space in all three directions, the cell count
is $640\times960\times270$, and the cell aspect ratio is
$\mathrelp\approx1:2:1$. The simulation formally terminates at $t=12\,t_*$,
where $t_*=2\pi a_*^{1/2}(a_*+2)$ is the orbital period at the characteristic
orbit, but we allow the simulation to run for $0.5\,t_*$ more so we can perform
time-averages centered around the end. This simulation duration is sufficient
for the \ac{MRI} to reach nonlinear saturation, but it is much too short for
the parametric instability of \citet{2005A&A...432..743P, 2005A&A...432..757P}
to appear.

The boundary conditions are outflow in the $\lambda$- and
$\zeta$\nobreakdash-directions and periodic in the
$\phi$\nobreakdash-direction. The $\lambda$\nobreakdash-boundaries copy all
quantities to the ghost zone, zero the $\lambda$\nobreakdash-component of the
velocity if it points into the simulation domain, and zero the $\phi$- and
$\zeta$\nobreakdash-components of the projected magnetic field always. The
$\zeta$\nobreakdash-boundaries are implemented analogously. Although the
magnetic field is not strictly divergence-free in the ghost zone, the finite
divergence does not propagate into the simulation domain.

Lastly, when the recovery of primitive variables fails, we carry over primitive
variables from the previous time step.

\section{Results}
\label{sec:results}

\subsection{Summary}
\label{sec:summary}

In terms of global gas properties, the stratified disk behaves much like the
unstratified eccentric disk in \citet{2022ApJ...933...81C} with a dipolar
magnetic topology: Gas spreads both inward and outward over the course of the
simulation, as we can see in \cref{fig:density}.

During this time, the stratified disk as a whole remains eccentric, but
eccentricity rises at small radii and falls at large radii as \ac{MHD} stresses
transport angular momentum outward. The inner parts of the disk undergo
prograde apsidal precession during $t\lesssim2\,t_*$, but this precession then
ceases.

\begin{figure}
\includegraphics{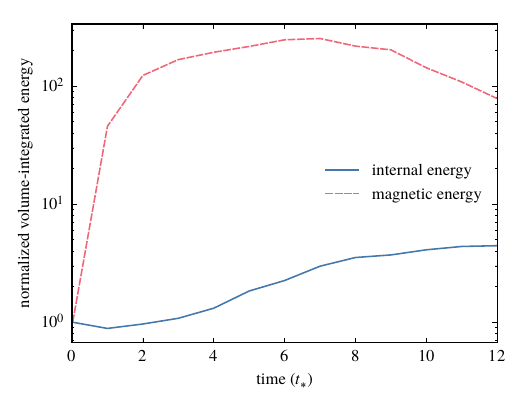}
\caption{Volume-integrated internal and magnetic energies, divided by
volume-integrated density to account for mass loss from the simulation domain.
Each curve is normalized to unity at $t=0$ to emphasize variation over time.}
\label{fig:energy}
\end{figure}

An important difference between the stratified disk and our earlier
unstratified one is that the stratified disk puffs up over time. In the
stratified case, vertical expansion is due to the adiabatic equation of state
retaining the internal energy created by dissipation of orbital energy.
\Cref{fig:energy} shows an increase in internal energy by a factor of
\numrange{\sim4}{5} by the end of the simulation. In the unstratified case, the
isothermal equation of state prevents any such internal energy retention, and
the absence of vertical gravity means vertical hydrostatic balance is
irrelevant in any event.

\begin{figure}
\includegraphics{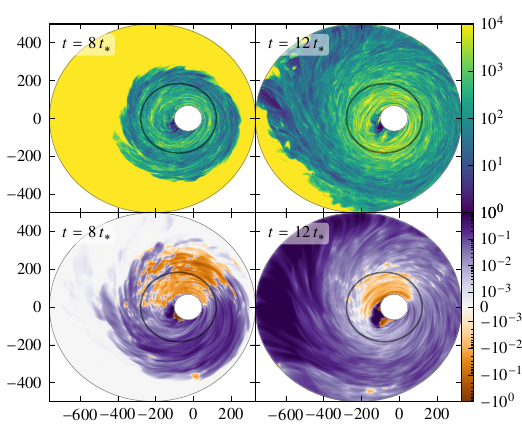}
\caption{Mass-weighted vertical averages of plasma beta in the top row and
Maxwell-only alpha parameter in the bottom row. The boundaries of the
simulation domain are traced by thin gray curves and the characteristic orbit
$\lambda=\lambda_*$ is indicated by a faint black ellipse.}
\label{fig:magnetic map}
\end{figure}

In terms of magnetic properties, we consider the ratio $\alpha_{\su m}$ of
Maxwell stress to pressure, which is the \citet{1973A&A....24..337S} $\alpha$
parameter but ignoring for simplicity the contribution from the Reynolds
stress. The stratified disk resembles its unstratified counterpart in that
$\alpha_{\su m}$, portrayed in the bottom row of \cref{fig:magnetic map}, is
consistently positive in one half of the disk and negative in the other half.

\Cref{fig:energy} charts how the magnetic energy per mass surges to a peak
value \numrange{\sim200}{300} times the initial value during the first half of
the simulation, then drops by a factor of \numrange{\sim3}{4} during the second
half. Weaker magnetization at late times is accompanied by a smaller
$\alpha_{\su m}$. This magnetic field decay may be partly due to reconnection
driven by compression resulting from the strong vertical gravity near
pericenter.

\subsection{Gas behavior}
\label{sec:gas}

\subsubsection{Radial and vertical expansion}

As \cref{fig:density} makes clear, the disk expands dramatically both radially
and vertically over the 12 orbits of the simulation. One way to quantify this
spreading is to note that, at $t=0$, \qty{\approx86}{\percent} of the mass is
contained within $\abs{\ln(\lambda/\lambda_*)}\le0.2$, or
$138\lesssim\lambda\lesssim205$, whereas the same fraction of the mass at
$t=12\,t_*$ is found within $\abs{\ln(\lambda/\lambda_*)}\le0.53$, or
$99\lesssim\lambda\lesssim285$. In other words, the disk is roughly two to
three times wider both inward and outward.

The vertical expansion seen in the middle rows of \cref{fig:density} is the
product of internal energy accumulation: \Cref{fig:energy} demonstrates that
the ratio of volume-integrated gas pressure to volume-integrated density of the
stratified disk rises steadily at $t\gtrsim3\,t_*$, albeit at slower rates
toward the end. For the unstratified disk with a dipolar magnetic topology, the
same diagnostic scarcely changes at all, as expected from the use of an
isothermal equation of state.

\subsubsection{Eccentricity and orientation}
\label{sec:eccentricity and orientation}

\begin{figure}
\includegraphics{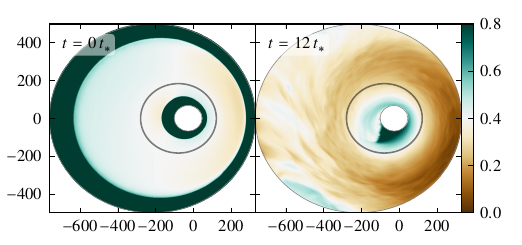}
\caption{Mass-weighted vertical averages of instantaneous eccentricity. The
boundaries of the simulation domain are traced by thin gray curves and the
characteristic orbit $\lambda=\lambda_*$ is indicated by a faint black
ellipse.}
\label{fig:eccentricity}
\end{figure}

To quantify eccentricity evolution, we compute the instantaneous eccentricity
$\bar e$, which is the eccentricity of the orbit a gas packet would follow
given its velocity if there were no forces other than gravitational. The accent
distinguishes this quantity from $e$ of the coordinate system. We use a
definition similar to \citet{2022ApJ...933...81C}, but to account for the
effect of vertical gravity, we first project the velocity to the midplane
before calculating $\bar e$; \cref{sec:eccentricity} describes how this is
done. \Cref{fig:eccentricity} depicts the mass-weighted vertical average of the
instantaneous eccentricity:
\begin{equation}
\mean{\bar e}_{\zeta;\rho}\eqdef
  \int d\zeta\,(-g)^{1/2}\rho\bar e\bigg/
  \int d\zeta\,(-g)^{1/2}\rho.
\end{equation}
Henceforth when mentioning averages like this, we drop for brevity the
subscript summarizing how the average is performed. At the start of the
simulation, $\mean{\bar e}\approx0.4$ over the entire disk. Small deviations
indicate non-Keplerian rotation, present because the circular disk on which the
initial eccentric disk is based features non-Keplerian rotation due to pressure
gradients. At the end of the simulation, $0.3\lesssim\mean{\bar e}\lesssim0.4$
near the characteristic orbit, $0.3\lesssim\mean{\bar e}\lesssim0.5$ inside it,
and $0.2\lesssim\mean{\bar e}\lesssim0.4$ outside. The most extreme $\mean{\bar
e}$ is found in regions farthest away from the characteristic orbit, both
inside and outside, but the great majority of these regions have very low
density according to \cref{fig:density}.

\begin{figure}
\includegraphics{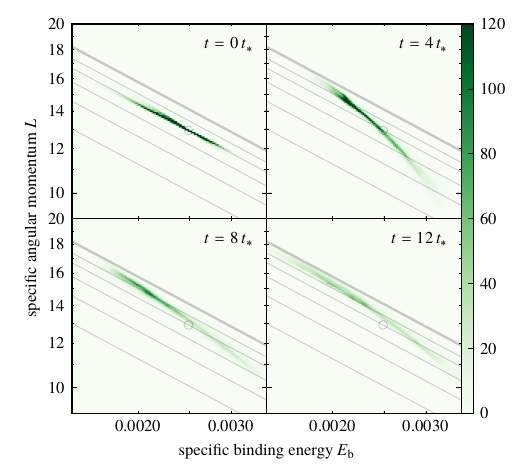}
\caption{Mass-weighted two-dimensional histograms of specific binding energy
$E_{\su b}$ and specific angular momentum $L$. The histograms are normalized by
the initial mass and the axes are logarithmic. The small faint circle marks the
$E_{\su b}$ and $L$ of the characteristic orbit $\lambda=\lambda_*$. The faint
lines trace, in order from top to bottom, the combinations of $E_{\su b}$ and
$L$ for which $\bar e\in\{0,0.3,0.4,0.5,0.6,0.7\}$; the histograms are
practically zero above the thick $\bar e=0$ line.}
\label{fig:energy and angular momentum}
\end{figure}

\Cref{fig:energy and angular momentum} depicts the mass distribution over
specific binding energy $E_{\su b}=1-E$ and specific angular momentum $L$,
where $E$ is the specific energy inclusive of rest energy. At all times, $\ln
L$ is linearly correlated with $\ln E_{\su b}$ to a good approximation.
\Ac{MHD} stresses transfer $L$ from the inner parts of the disk with lower
initial $L$ to the outer parts with higher initial $L$. Compared to the large
fractional change in $L^2$, the fractional change in $E_{\su b}$ is much
smaller. The net result is that the slope of the correlation steepens from an
initial value of $-\tfrac12$ to \num{\sim-0.6} at $t=12\,t_*$, with a slightly
steeper slope for low\nobreakdash-$L$ material. Because $E_{\su
b}L^2\approx\tfrac12(1-\bar e^2)$ if $E_{\su b}\ll1$, eccentricity rises in the
inner parts and falls in the outer parts.

\begin{figure}
\includegraphics{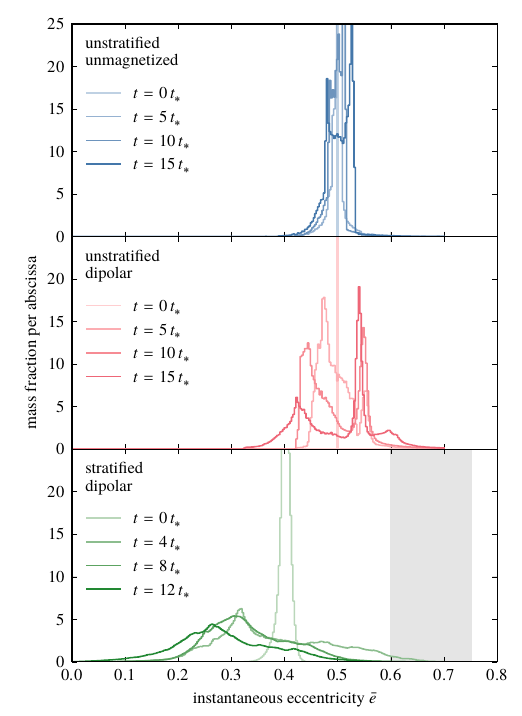}
\caption{Mass-weighted histograms of instantaneous eccentricity for several
eccentric disks. The top and middle panels are for two unstratified disks from
\citet{2022ApJ...933...81C}, the top panel for the unmagnetized disk and the
middle panel for the disk with a dipolar magnetic topology. The bottom panel is
for the stratified disk from this article, which also has a dipolar magnetic
topology; the gray band approximately indicates the $\bar e$ of the material
crossing the inner boundary.}
\label{fig:eccentricity comparison}
\end{figure}

Another view of the eccentricity evolution is provided by
\cref{fig:eccentricity comparison}, which compares the mass distribution over
$\bar e$ of the stratified disk with two unstratified disks from
\citet{2022ApJ...933...81C}. All three disks have narrow initial distributions
that spread over time; the spread is very slow for the unstratified
unmagnetized disk and fastest for the stratified disk. The unstratified
distributions shown acquire a bimodal shape centered on the initial $\bar e$.
By contrast, the stratified distribution remains unimodal and shifts toward
smaller $\bar e$. We will discuss these contrasting evolutions in
\cref{sec:rapid evolution,sec:transience}. The apparent asymmetry of the
stratified distribution is partly because material with $0.60\lesssim\bar
e\lesssim0.75$ has such small pericentric distances that it is lost through the
inner boundary; \qty{\sim9}{\percent} of the initial mass is lost in this way.

The extent to which the stratified disk undergoes apsidal precession can be
determined by tracking the azimuth of the pericenter, which is also the
thinnest part of the disk. \Cref{fig:density} shows the inner parts of the disk
undergo reorientation: They precess by $\mathrelp\sim\tfrac16\pi$ in the
prograde direction within $t\lesssim2\,t_*$ and stay at roughly that
orientation for the rest of the simulation. In comparison, the outer parts do
not alter their orientation appreciably. Similar to the unstratified disks in
\citet{2022ApJ...933...81C}, apsidal precession in the stratified disk is
limited both in duration and extent; hence, its long-term effect is likely
small.

The transient apsidal precession in our stratified and unstratified simulations
reflects the relaxation of the disk from its initial condition. It should be
distinguished from the secular pressure-induced apsidal precession discussed in
other contexts of eccentric disks \citep[e.g.,][]{2001AJ....122.2257S}. The
rate of such apsidal precession is of order the orbital frequency divided by
the square of the Mach number \citep[e.g.,][]{2000ssd..book.....M}, which is
much slower than the apsidal precession we witness.

\subsubsection{Disk height}

The initial disk breathing matches grid breathing perfectly by construction,
but the disk gradually drifts away from that perfect match over time. Apart
from apsidal precession, we see other signs of mismatch by the end of the
simulation in the middle rows of \cref{fig:density}: The disk height near
apocenter rises by a factor of \num{\sim2.5} over the course of the simulation
but hardly changes near pericenter, and the density distribution and disk
height are no longer symmetric about the line of apsides. The larger
apocenter-to-pericenter disk height ratio than grid height ratio was
anticipated in \cref{sec:coordinate system}.

\begin{figure}
\includegraphics{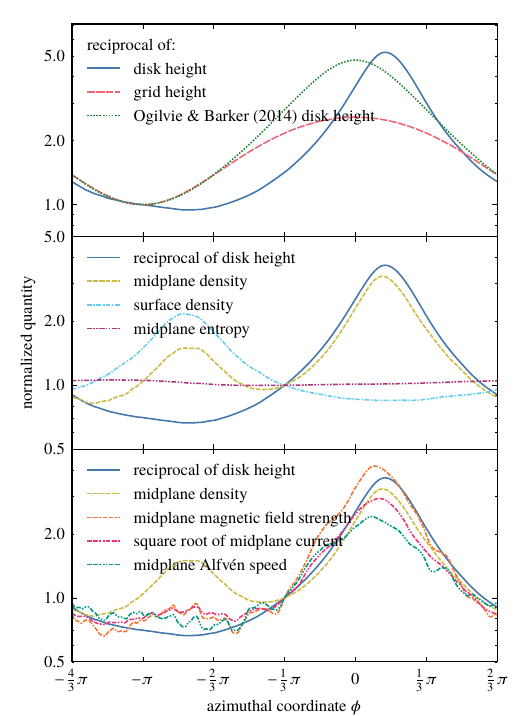}
\caption{Azimuthal modulation along the characteristic orbit
$\lambda=\lambda_*$ at $t=12\,t_*$; see \cref{sec:gas,sec:magnetic} for
definitions. The abscissa is shifted so that the apocentric region fits in the
left half of each panel and the pericentric region in the right half. To
emphasize relative change, the curves in the top, middle, and bottom panels are
normalized by their values at the arbitrarily chosen azimuths $\phi=-\pi$,
$\phi=-\tfrac13\pi$, and $\phi=-\tfrac13\pi$, respectively.}
\label{fig:modulation}
\end{figure}

For a more quantitative examination of the disk height modulation, we calculate
the mass-weighted distance from the midplane:
\begin{multline}\label{eq:disk height}
\mean{\abs z}_{t\lambda\zeta;\rho}\eqdef
  \int dt\,d\ln\lambda\,d\zeta\,(-g)^{1/2}\rho\abs z\bigg/ \\
  \int dt\,d\ln\lambda\,d\zeta\,(-g)^{1/2}\rho.
\end{multline}
The temporal integration limits are $11.5\,t_*\le t\le12.5\,t_*$;
time-averaging is done in consistency with the magnetic quantities discussed in
\cref{sec:magnetic}, which require time-averaging to smooth out strong spatial
fluctuations. The spatial integration limits are
$-0.2\le\ln(\lambda/\lambda_*)\le0.2$ and all $\zeta$; although the disk
evolves in eccentricity and orientation, we stick to an elliptical annulus
around the characteristic orbit for simplicity. The top panel of
\cref{fig:modulation} traces $\mean{\abs z}$, or rather its reciprocal, in
anticipation of how other quantities in the same figure behave. We also show
the reciprocal $H_*^{-1}$ of the grid height for context. The modulation of
$\mean{\abs z}$ around the orbit is stronger than $H_*$. The extrema of
$\mean{\abs z}$ are displaced in azimuth by $\mathrelp\sim\tfrac16\pi$ from
those of $H_*$, reflecting apsidal precession.

\Citet{2014MNRAS.445.2621O} wrote down the equation governing the height of a
laminar eccentric disk that expands and contracts homologously in the vertical
direction. We reproduce in the same panel the reciprocal of the solution to
their equation for a uniformly eccentric disk with the same eccentricity and
adiabatic index as our initial disk. The analytic disk height varies with an
amplitude close to our $\mean{\abs z}$, but the rise and fall of our
$\mean{\abs z}$ near pericenter is more rapid.

\subsubsection{Azimuthal modulation}

Next we inspect the azimuthal modulation of various hydrodynamic quantities,
beginning with density $\rho$ and surface density
\begin{equation}
\Sigma\eqdef R^{-2}\int d\zeta\,(-g)^{1/2}\rho
\end{equation}
as shown in \cref{fig:density}. The initial surface density is uniform over
azimuth; hence, the initial density peaks at the pericenter $\phi=0$. Apsidal
precession moves the pericenter and the density peak to $\phi\sim\tfrac16\pi$;
at the same time, the surface density develops a maximum at
$\phi\sim-\tfrac56\pi$ near apocenter.

To put this on a more quantitative footing, we construct the azimuthal profiles
of midplane density
\begin{equation}\label{eq:density}
\mean\rho_{t\lambda\zeta;}\eqdef
  \int dt\,d\ln\lambda\,d\zeta\,(-g)^{1/2}\rho\bigg/
  \int dt\,d\ln\lambda\,d\zeta\,(-g)^{1/2}
\end{equation}
and surface density
\begin{equation}
\mean\Sigma_{t\lambda;R^2}\eqdef
  \int dt\,d\ln\lambda\,d\zeta\,(-g)^{1/2}\rho\bigg/
  \int dt\,d\ln\lambda\,R^2.
\end{equation}
The integration limits for $\mean\Sigma$ are the same as for $\mean{\abs z}$ in
\cref{eq:disk height}, and those for $\mean\rho$ differ only in that
$-0.2\le\zeta\le0.2$, to pick out the midplane. The results are displayed in
the middle panel of \cref{fig:modulation}, along with the reciprocal
$\mean{\abs z}^{-1}$ of the disk height. The secondary peak of $\mean\rho$ at
$\phi\sim-\tfrac56\pi$ is barely noticeable in \cref{fig:density}. Over
$-\tfrac13\pi\lesssim\phi\lesssim\tfrac23\pi$ around pericenter, we have
$\mean\rho\mean{\abs z}\propto\mean\Sigma$ and constant $\mean\Sigma$,
consistent with pure vertical compression; over
$-\tfrac43\pi\lesssim\phi\lesssim-\tfrac13\pi$ around apocenter, we still have
$\mean\rho\mean{\abs z}\propto\mean\Sigma$, but the non-constancy of
$\mean\Sigma$ implies that, unlike in the initial disk, horizontal motion is
not divergence-free.

Following up on the increase in internal energy due to dissipation shown in
\cref{fig:energy}, we graph in the same panel of \cref{fig:modulation} a proxy
for the mass-weighted midplane entropy,
\begin{multline}
\mean s_{t\lambda\zeta;\rho}\eqdef
  \int dt\,d\ln\lambda\,d\zeta\,(-g)^{1/2}\rho
    \ln\biggl(\frac p{\rho^\gamma}\biggr)\bigg/ \\
  \int dt\,d\ln\lambda\,d\zeta\,(-g)^{1/2}\rho,
\end{multline}
reusing here the integration limits for $\mean\rho$ in \cref{eq:density}. The
flatness of $\mean s$ indicates that, in this simulation with heating but no
cooling, entropy growth is slow compared to orbital motion by the end of the
simulation.

\subsection{Magnetic behavior}
\label{sec:magnetic}

\subsubsection{Magnetic field strength and Maxwell stress}
\label{sec:magnetic field and stress}

The magnetic field grows in a way similar to circular disks. The initial
magnetic field is symmetric about the midplane. As the simulation progresses,
differential rotation draws out the magnetic field, the \ac{MRI} grows in
amplitude from small to nonlinear, parasitic instabilities break up the smooth
azimuthal variation of the magnetic field \citep[e.g.,][]{1994ApJ...432..213G},
and full three-dimensional turbulence begins to grow. The symmetry about the
midplane lasts until $t\sim6\,t_*$, shortly after the ordered stage of magnetic
field growth ends, but shortly before the volume-integrated magnetic energy
reaches a maximum at $t\sim7\,t_*$, according to \cref{fig:energy}. Thereafter,
the magnetic field becomes increasingly turbulent even while the
volume-integrated magnetic energy decays.

The bottom panel of \cref{fig:modulation} traces the azimuthal profile of the
midplane magnetic field strength
\begin{multline}
\mean b_{t\lambda\zeta;}\eqdef\biggl[
  \int dt\,d\ln\lambda\,d\zeta\,(-g)^{1/2}b_\mu b^\mu\bigg/ \\
  \int dt\,d\ln\lambda\,d\zeta\,(-g)^{1/2}
\biggr]^{1/2}.
\end{multline}
The integration limits are the same as $\mean\rho$ in \cref{eq:density}. Over
$-\tfrac13\pi\lesssim\phi\lesssim\tfrac23\pi$ around pericenter, vertical
compression and flux freezing explain why $\mean b\propto\mean{\abs z}^{-1}$.
The symmetric rise and fall of $\mean b$ during pericenter passage suggest that
the magnetic field is not noticeably amplified by effects other than
compression. This nonlinear-stage behavior is different from the pericentric
amplification seen in linear-stage \ac{MRI} \citep{2018ApJ...856...12C}.

Another way to examine the magnetic field strength is through the mass-weighted
vertical average of the plasma beta:
\begin{equation}
\mean\beta_{\zeta;\rho}\eqdef
  \int d\zeta\,(-g)^{1/2}\rho\frac{2p}{b_\mu b^\mu}\bigg/
  \int d\zeta\,(-g)^{1/2}\rho.
\end{equation}
The top row of \cref{fig:magnetic map} shows that
$\num{e2}\lesssim\mean\beta\lesssim\num{e4}$. The value of $\mean\beta$
increases over time, in agreement with the magnetic field decay noted above. It
is not a strong function of azimuth because vertical compression enhances gas
and magnetic pressures by similar amounts.

To compute the Maxwell stress in cylindrical coordinates, we need the physical,
cylindrical components $b^{\hat i}$ of the projected magnetic field, that is,
the components measured in a local orthonormal basis whose basis vectors are
parallel to those of cylindrical coordinates. They are related to the
contravariant components $b^i$ by
\begin{align}
\label{eq:cylindrical magnetic field 1}
b^{\hat R} &= R\biggl(b^\lambda+b^\phi\frac{e\sin\phi}{1+e\cos\phi}\biggr), \\
\label{eq:cylindrical magnetic field 2}
b^{\hat\varphi} &= Rb^\phi, \\
\label{eq:cylindrical magnetic field 3}
b^{\hat z} &=
  z\biggl(b^\lambda+b^\phi\od{}\phi\ln H_*+\frac{b^\zeta}\zeta\biggr),
\end{align}
and they satisfy the normalization
\begin{equation}
g_{ij}b^ib^j=
  b^{\hat R}b^{\hat R}+b^{\hat\varphi}b^{\hat\varphi}+b^{\hat z}b^{\hat z}.
\end{equation}
Following \citet{2022ApJ...933...81C}, we ignore the contribution of the
Reynolds stress to the total internal stress on the assumption that the Maxwell
stress dominates, an extrapolation from circular disks
\citep[e.g.,][]{1995ApJ...440..742H}. The Maxwell-only alpha parameter is
\begin{equation}
\alpha_{\su m}=-b^{\hat\varphi}b^{\hat R}/p.
\end{equation}

The bottom row of \cref{fig:magnetic map} depicts the vertical averages of
$\alpha_{\su m}$ at $t=8\,t_*$ and $t=12\,t_*$:
\begin{equation}
\mean{\alpha_{\su m}}_{\zeta;p}\eqdef
  -\int d\zeta\,(-g)^{1/2}b^{\hat\varphi}b^{\hat R}\bigg/
   \int d\zeta\,(-g)^{1/2}p.
\end{equation}
The magnitude of $\mean{\alpha_{\su m}}$ at both times is
$\num{e-3}\lesssim\abs{\mean{\alpha_{\su m}}}\lesssim\num{e-1}$, roughly
comparable to the values seen in global simulations of circular disks
\citep[e.g.,][]{2000ApJ...528..462H, 2002ApJ...566..164H}. However, the
negative sign of $\mean{\alpha_{\su m}}$ in the half of the disk where material
flows outward from pericenter to apocenter is never seen in circular disks, but
is found in unstratified eccentric disks \citep{2022ApJ...933...81C}. We
therefore identify this sign flip with the same mechanism responsible for it in
the unstratified case, that is, the change in sign of $\pds{u^{\hat
R}}\varphi$, which alters the correlation of $b^{\hat R}$ and $b^{\hat\varphi}$
created by orbital shear. Consistent with this interpretation, at $t=8\,t_*$
the $\mean{\alpha_{\su m}}<0$ region stretches across the entire radial extent
of the disk, whereas at $t=12\,t_*$ when the outer parts of the disk are
substantially less eccentric, the $\mean{\alpha_{\su m}}<0$ region is more
limited in size.

The same figure shows that the magnitude of $\mean{\alpha_{\su m}}$ declines
over time: $\num{3e-3}\lesssim\abs{\mean{\alpha_{\su m}}}\lesssim\num{3e-1}$ at
$t=8\,t_*$ but $\num{e-3}\lesssim\abs{\mean{\alpha_{\su m}}}\lesssim\num{e-1}$
at $t=12\,t_*$. We associate the late-time decline with a general weakening of
the magnetic field after $t\gtrsim7\,t_*$, as illustrated in \cref{fig:energy}.
This effect may be attributable to the strong reconnection in stratified
eccentric disks, an idea we explore further in \cref{sec:reconnection}.

Another way to average $\alpha_{\su m}$ is over constant\nobreakdash-$R$
cylinders, taking into account the
\qtyrange[range-units=repeat]{10}{20}{\percent} contribution of $\alpha_{\su
m}<0$ regions to the average. On cylinders running through denser parts of the
disk, this average rises rapidly to \num{\sim0.1} at $t\sim2\,t_*$, falls to
\num{\sim0.05} at $t\sim5\,t_*$ as the \ac{MRI} turns nonlinear, and dwindles
to \num{\sim0.01} at $t\sim12\,t_*$ with the gradual diminution of the magnetic
field. At $t\gtrsim8\,t_*$, $\alpha_{\su m}$ is inversely correlated with $\bar
e$: $\alpha_{\su m}$ averaged on cylinders at $R\sim300$, where orbits are less
eccentric, is a few times that on cylinders at $R\sim200$.

\subsubsection{Reconnection}
\label{sec:reconnection}

Reconnection is the principal mechanism for magnetic field decay. Because
\acp{MHD} simulations do not resolve the microscopic scales on which physical
reconnection occurs, reconnection in a simulation like ours is necessarily
numerical: Oppositely directed magnetic fields advected into the same cell
cancel each other. Reconnection in our simulation happens in regions of strong
magnetic shear, a measure of which is $\nabla_\nu F^{\mu\nu}=j^\mu$; its
magnitude $(j_\mu j^\mu)^{1/2}$ is the magnitude of the usual three-current in
the comoving frame. Studies of magnetic dissipation in grid-based simulations
found that the dissipation rate scales with the current magnitude slightly
faster than linearly \citep{2006ApJ...640..901H}.

\begin{figure}
\includegraphics{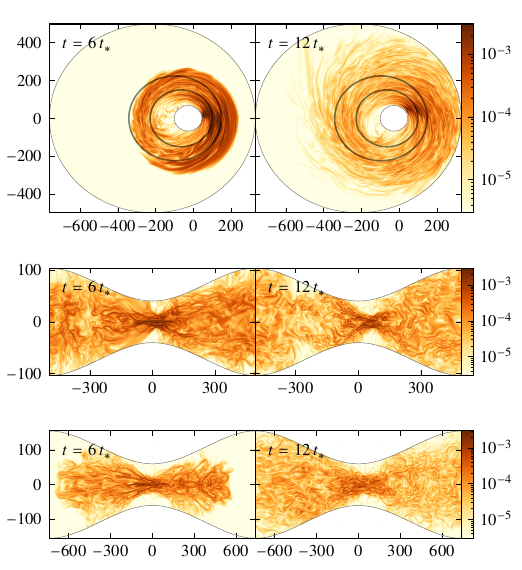}
\caption{Magnitude of current. The top row shows midplane slices, and the
middle and bottom rows show vertical slices along
$\ln(\lambda/\lambda_*)=\mp0.2$, respectively, on either side of the
characteristic orbit. The boundaries of the simulation domain are traced by
thin gray curves. For the top row, the ellipses along which vertical slices are
taken are indicated by faint black ellipses. For the middle and bottom rows,
the abscissa is the physical arc length along the orbit measured from $\phi=0$,
the ordinate is the physical height $z$ above the midplane, and the ordinate is
more stretched than the abscissa.}
\label{fig:current}
\end{figure}

It is therefore instructive to plot $(j_\mu j^\mu)^{1/2}$, as in
\cref{fig:current}. The current is strongly enhanced near pericenter because
pericentric compression both strengthens the magnetic field and sharpens its
gradient. To examine its azimuthal variation more quantitatively, we consider
the midplane current magnitude
\begin{multline}
\mean j_{t\lambda\zeta;}\eqdef\biggl[
  \int dt\,d\ln\lambda\,d\zeta\,(-g)^{1/2}j_\mu j^\mu\bigg/ \\
  \int dt\,d\ln\lambda\,d\zeta\,(-g)^{1/2}
\biggr]^{1/2},
\end{multline}
the integration limits being the same as $\mean\rho$ in \cref{eq:density}. The
bottom panel of \cref{fig:modulation} shows that $\mean j\propto\mean{\abs
z}^{-2}$. This scaling results from the magnetic field strength growing as
$\mathrelp\propto\mean{\abs z}^{-1}$ when toroidal magnetic field is compressed
vertically and the length scale of the vertical gradient varying as
$\mathrelp\propto\mean{\abs z}$; the current is proportional to the ratio of
these two quantities.

\subsubsection{Quality factors}
\label{sec:quality}

\begin{figure*}
\includegraphics{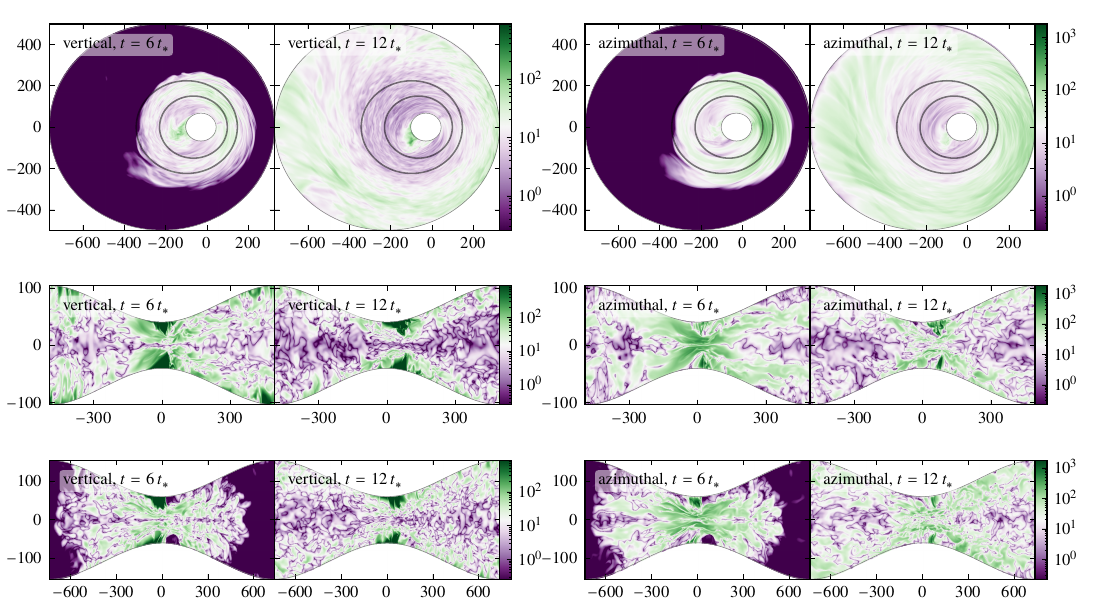}
\caption{Vertical and azimuthal quality factors. The top row shows
mass-weighted vertical averages, and the middle and bottom rows show vertical
slices along $\ln(\lambda/\lambda_*)=\mp0.2$, respectively, on either side of
the characteristic orbit. The boundaries of the simulation domain are traced by
thin gray curves. For the top row, the ellipses along which vertical slices are
taken are indicated by faint black ellipses. For the middle and bottom rows,
the abscissa is the physical arc length along the orbit measured from $\phi=0$,
the ordinate is the physical height $z$ above the midplane, and the ordinate is
more stretched than the abscissa.}
\label{fig:quality}
\end{figure*}

The degree to which the \ac{MRI} is resolved is assessed by quality factors.
The forms adopted here are simple generalizations from
\citet{2011ApJ...738...84H} and are almost identical to
\citet{2022ApJ...933...81C}:
\begin{align}
Q_z &\eqdef \frac{\abs{b^{\hat z}}t_{\su o}}
  {\rho^{1/2}g_{\zeta\zeta}^{1/2}\Delta\zeta}, \\
Q_\varphi &\eqdef \frac{\abs{b^{\hat\varphi}}t_{\su o}}
  {\rho^{1/2}R\Delta\phi},
\end{align}
where $t_{\su o}=2\pi a^{1/2}(a+2)$, with $a=\lambda/(1-e^2)$, is the orbital
period at $\lambda$. \Cref{fig:quality} visualizes the quality factors at
$t=6\,t_*$, when the magnetic field is strongest, and at $t=12\,t_*$, at the
end of the simulation. Mass-weighted vertical averages appear in the top row:
\begin{align}
\mean{Q_z}_{\zeta;\rho} &\eqdef
  \int d\zeta\,(-g)^{1/2}\rho Q_z\bigg/
  \int d\zeta\,(-g)^{1/2}\rho, \\
\mean{Q_\varphi}_{\zeta;\rho} &\eqdef
  \int d\zeta\,(-g)^{1/2}\rho Q_\varphi\bigg/
  \int d\zeta\,(-g)^{1/2}\rho;
\end{align}
raw $Q_z$ and $Q_\varphi$ on vertical slices appear in the middle and bottom
rows. The color scales are centered on $Q_z\sim15$ and $Q_\varphi\sim20$,
reflecting the criteria for adequate resolution in circular disks
\citep{2013ApJ...772..102H}.

There is clear azimuthal dependence of $Q_z$ and $Q_\varphi$, with both being
much higher near pericenter. There is also clear vertical dependence of $Q_z$
and $Q_\varphi$, with both being lower near the midplane where the gas is
denser and the magnetic field weaker. The former dependence can be understood
by examining the azimuthal variations of $v_{\su A}$, $H_*$, and $R$ along an
orbit. The bottom panel of \cref{fig:modulation} tells us that the
mass-weighted midplane Alfv\'en speed
\begin{multline}
\mean{v_{\su A}}_{t\lambda\zeta;\rho}\eqdef
  \int dt\,d\ln\lambda\,d\zeta\,(-g)^{1/2}\rho
    \biggl(\frac{b_{\smash\mu}b^\mu}\rho\biggr)^{1/2}\bigg/ \\
  \int dt\,d\ln\lambda\,d\zeta\,(-g)^{1/2}\rho,
\end{multline}
where the integration limits are the same as $\mean\rho$ in \cref{eq:density},
is a factor of \num{\sim3} higher near pericenter than near apocenter; the fact
that $\mean{v_{\su A}}$ reaches a maximum at pericenter is consistent with our
earlier observations that $\mean\rho\propto\mean{\abs z}^{-1}$ and $\mean
b\propto\mean{\abs z}^{-1}$. We see from the top panel of the same figure that
$H_*$ is a factor of \num{\approx2.6} lower near pericenter, and we can also
convince ourselves using elementary geometry that $R$ is a factor of
$(1+e)/(1-e)\approx2.3$ lower. It follows that both $Q_z\propto v_{\su A}/H_*$
and $Q_\varphi\propto v_{\su A}/R$ are maximum near pericenter, meaning that
the \ac{MRI} is best resolved there. This is in marked contrast to
hydrodynamics, which is harder to resolve near pericenter where the disk is
thinnest.

The quality factors are much better at $t=6\,t_*$ than $t=12\,t_*$. Our
extremely high resolution is able to provide at $t=6\,t_*$ tolerable quality
factors according to standards based on the circular \ac{MRI}
\citep{2013ApJ...772..102H}. The situation changes by $t=12\,t_*$, when the
quality factors have dropped by factors of \numrange{\sim2}{3}. Unfortunately,
an order-unity change in resolution quality can have strong effects on
simulation quality \citep{2013ApJ...772..102H}.

The question arises whether the factor of \num{\sim3} decline in magnetic
energy per mass in \cref{fig:energy} is physical or numerical; that is, how
much of it is due to strong magnetic dissipation near pericenter and how much
is due to poor resolution suppressing the \ac{MRI}. A weaker magnetic field, no
matter the cause, demands smaller cells to resolve both the linear and
nonlinear stages of the \ac{MRI}, and such deterioration in resolution may
precipitate further magnetic field decay. Because our resolution more than
suffices to follow \ac{MRI}-driven turbulence at $t=6\,t_*$, we believe at
least the initial decline is physical. However, the subsequent factor of
\num{\sim3} decrease in magnetic energy per mass means the quality factors
$\mathrelp\propto(B^2/\rho)^{1/2}$ at late times are, by definition, worse by
factors of unity. The poorer resolution slows down perturbation growth and
speeds up grid-scale dissipation, leading to magnetic field decay. The
reduction in magnetic field strength in this stage thus has both physical and
numerical aspects.

It is equally possible that the eccentric \ac{MRI} may put somewhat higher
demands on resolution because eccentric disks support linear \ac{MRI} growth in
modes with shorter wavelengths than would be possible in circular disks
characterized by the same product of Alfv\'en speed and orbital period
\citep{2018ApJ...856...12C}.

\section{Discussion}
\label{sec:discussion}

\subsection{\Acs*{MRI} and \acs*{MHD} stresses in eccentric disks}

This is the third in a series of articles investigating whether the \ac{MRI},
which has been studied extensively in circular disks, remains an instability
when orbits are eccentric, and if it does, to what degree its nonlinear,
saturated state resembles that found when orbits are circular. In our first
article \citep{2018ApJ...856...12C}, we verified that the \ac{MRI} remains
linearly unstable in unstratified eccentric disks; its growth rate differs only
by factors of unity from the circular case, but with increasing eccentricity,
the range of unstable vertical wavenumbers extends well beyond the maximum in
circular disks. In our second article \citep{2022ApJ...933...81C}, we showed
that unstratified circular and eccentric disks have comparable ratios of
Maxwell stress to pressure: $\alpha_{\su m}\sim\num{e-2}$ when averaged on
cylinders.

In this third article, we present the first calculation of the nonlinear
evolution of the \ac{MRI} in an eccentric disk with vertical gravity.
\Cref{sec:magnetic field and stress} showcases the similarity between
unstratified and stratified eccentric disks: Vertically averaged $\alpha_{\su
m}$ has positive and negative regions, and cylindrically averaged $\alpha_{\su
m}$ reaches $\num{\sim e-1}$ within a few orbits. However, in the presence of
vertical gravity, cylindrically averaged $\alpha_{\su m}$ eventually decays to
$\num{\sim e-2}$ by the end of the simulation; \cref{sec:quality} discusses to
what degree this diminution is physical or numerical. Nonetheless, our fully
three-dimensional simulation, incorporating vertical gravity, confirms that the
\ac{MRI} is an inescapable feature of eccentric disks, just as it is for
circular disks.

\subsection{Enhanced reconnection due to vertical compression}

\Ac{MHD} turbulence in circular disks leads to dissipation, much of which is
likely due to reconnection facilitated by the mechanism suggested by
\citet{1999ApJ...517..700L}. Our simulation uncovers a new channel specific to
eccentric disks that supplements and perhaps expedites turbulent dissipation
\citep[see also the discussion by][]{2021MNRAS.501.5500L}.

Circular disks feature no regions of coherent compression around the orbit. By
contrast, both unstratified and stratified eccentric disks experience
horizontal compression near pericenter thanks to the eccentric nature of the
orbits. On top of that, stratified eccentric disks undergo vertical compression
because of the strong vertical gravity near pericenter. Vertical gravity
strengthens rapidly with decreasing distance from the central object, becoming
$[(1+\bar e)/(1-\bar e)]^3$ times stronger at pericenter than at apocenter;
therefore, vertical compression can be stronger than horizontal even at
moderate eccentricities. When the relative compression speed between
neighboring gas packets is supersonic, shocks form that dissipate kinetic
energy \citep{2021ApJ...920..130R}.

In magnetized eccentric disks, horizontal and vertical compression may lead to
additional dissipation as a result of reconnection. This dissipation can occur
whether the compression speed is subsonic or supersonic; all it requires is
that compression brings oppositely directed magnetic fields closer.

Our disk, with a dipolar magnetic topology, illustrates a condition
particularly conducive to reconnection driven by vertical compression. The
initial magnetic field, which is approximately radial, points in opposite
directions on either side of the midplane. From this radial magnetic field,
orbital shear creates an azimuthal magnetic field with the same direction
contrast. Converging vertical flows then produce regions of consistently
powerful currents near the midplane in the pericenter region, visible in
\cref{fig:modulation,fig:current}; these regions are sites of intense magnetic
dissipation in our simulation. We speculate that the additional dissipation due
to this compression-driven reconnection causes the magnetic energy in our
simulation to decline after reaching a maximum at $t\sim7\,t_*$. We further
speculate that at $t\gtrsim8\,t_*$, the reduction in eccentricity and
compression-driven dissipation at large radii leads to the increase in
cylindrically averaged $\alpha_{\su m}$ with $R$ noted in \cref{sec:magnetic
field and stress}.

Our argument for a boost in reconnection due to vertical compression rests on
the horizontal magnetic field reversing direction across the midplane. In
addition to our dipolar magnetic topology, this could also occur with a
toroidal magnetic field if adjacent field loops point in opposite directions.
Other magnetic topologies may not share this property. A toroidal magnetic
field may avoid accelerated reconnection if all of its field loops point in a
single direction. Likewise, a quadrupolar magnetic field with two poloidal
field loops above and below the midplane may not suffer elevated levels of
dissipation because the region of greatest compression around the midplane has
magnetic field running in the same direction.

\subsection{Rapidity of \acsp*{MHD}-driven evolution in eccentric disks}
\label{sec:rapid evolution}

Global \acp{MHD} simulations of circular disks generally exhibit an initial
transient stage of structural change on the orbital timescale, followed by an
extended period of gradual evolution on the inflow timescale. Two-dimensional,
purely hydrodynamical investigations of uniformly eccentric disks show that
pressure forces can reshape the disk on the secular timescale
\citep[e.g.,][]{2000ssd..book.....M}, but the disk can retain its eccentricity
and orientation distributions, precessing rigidly, if it starts in an eccentric
mode \citep[e.g.,][]{2001AJ....122.2257S}.

In sharp contrast to these outcomes, \cref{sec:eccentricity and orientation}
demonstrates that \ac{MHD} effects in an eccentric disk, whether stratified or
unstratified, can transform the radial profile of eccentricity $\bar e$ and
alter disk structure over just \numrange{\sim5}{10} orbits, as the contrast
between the top and middle panels of \cref{fig:eccentricity comparison} makes
clear. This may be surprising at first glance given that $\alpha_{\su m}$
indicates \ac{MHD} stresses comparable to the ones driving much slower
evolution in circular disks.

Two facts partially explain why. The first fact is that disk material largely
maintains its specific binding energy $E_{\su b}$ throughout the simulation, as
\cref{fig:energy and angular momentum} suggests. For fixed $E_{\su b}$, the
change in $\bar e$ due to the transfer of specific angular momentum $L$ is
given by $\pds{\ln(1-\bar e^2)}L=2/L\propto(1-\bar e^2)^{-1/2}E_{\su b}^{1/2}$
if $E_{\su b}\ll1$. Consequently, in moderately eccentric disks, a modest
amount of angular momentum loss can cause a large fractional increase in
$1-\bar e^2$ in parts where $\bar e$ is already large, as can be seen in
\cref{fig:energy and angular momentum,fig:eccentricity comparison}.

A second fact contributing to this result is that \ac{MHD} effects can catalyze
stronger hydrodynamic torques. Examination of the local hydrodynamic and
magnetic torques, defined as the divergence of the respective stress tensors,
reveals that hydrodynamic torques in the dipolar-field disks shown in
\cref{fig:eccentricity comparison} are generically two orders of magnitude
stronger than magnetic torques and a factor of a few stronger than hydrodynamic
torques in the unmagnetized disk. This may be the reason why the magnetized
disks in \cref{fig:eccentricity comparison} evolve much faster than the
unmagnetized disk.

\subsection{Transience of magnetized eccentric disks}
\label{sec:transience}

The character of the rapid evolution discussed in \cref{sec:rapid evolution} is
to broaden the mass distribution over specific angular momentum and steepen the
negative correlation between specific angular momentum $L$ and specific binding
energy $E_{\su b}$, making the disk in \cref{fig:eccentricity} more eccentric
on its inside and more circular on its outside. During our relatively brief
simulation duration of 12 orbits, there is no indication of either trend
slowing down, as \cref{fig:energy and angular momentum} and the bottom panel of
\cref{fig:eccentricity comparison} make apparent, or the disk approaching
anything resembling a steady state.

The evolution toward a structure more circular on the outside and more
eccentric on the inside is a result of the torque distribution around the
orbit. This result is most easily derived in the Newtonian limit of $E_{\su
b}\ll1$, in which the eccentricity $\bar e$ of an orbit changes according to
\begin{equation}
\frac{d(1-\bar e^2)}{1-\bar e^2}=2\frac{dL}L+\frac{dE_{\su o}}{E_{\su o}}
  =2\biggl(1-\frac{\Omega L}{2E_{\su b}}\biggr)\frac{dL}L,
\end{equation}
where $E_{\su o}\eqdef E-1=-E_{\su b}$ is the specific orbital energy, and we
used the fact that the work done $dE_{\su o}$ is the local orbital frequency
$\Omega$ times $dL$. The first and second terms describe, respectively, the
eccentricity change due to angular momentum and energy changes, and they have
opposite signs.

Because $L=[\bar a(1-\bar e^2)]^{1/2}$, $E_{\su b}=1/(2\bar a)$, and
$\mean\Omega=\bar a^{-3/2}$, with $\mean\Omega$ and $\bar a$ the mean motion
and semimajor axis of the orbit, respectively, this expression can be
fruitfully simplified to
\begin{equation}
\frac{d(1-\bar e^2)}{1-\bar e^2}
  =2\biggl[1-(1-\bar e^2)^{1/2}\frac\Omega{\mean\Omega}\biggr]\frac{dL}L.
\end{equation}
From this form, it can be immediately seen that whether the first or the second
term is greater depends on the ratio of local to mean orbital frequency, and
the term related to energy change is down-weighted when $\bar e$ is large.

Assuming orbital evolution is slow enough that $\bar e$, $L$, and $E_{\su b}$
are almost constant over an orbit, the change in $1-\bar e^2$ in a single orbit
is
\begin{multline}
\frac{\Delta(1-\bar e^2)}{1-\bar e^2}
\approx2
  \biggl[1-(1-\bar e^2)^{1/2}\frac{\int dL\,\Omega}{\Delta L\mean\Omega}\biggr]
  \frac{\Delta L}L \\
=2[1-(1-\bar e^2)^{1/2}]\frac{\Delta L}L,
\end{multline}
where the last step is true in the simplest case of constant $\ods Lt$
\citep[see also][]{2017MNRAS.467.1426S}. In this case, the change in
eccentricity due to torque is always at least as great as that due to work.
Whether the same conclusion holds if the torque is not constant depends on
whether more of $\Delta L$ is accomplished near apocenter, where $\Omega$ is
relatively small, or near pericenter, where it is relatively large; the
criterion for torque-dominance becomes easier to meet for larger $\bar e$.

Moreover, if the torque term dominates the work term, $\Delta\bar e$ has the
opposite sign to $\Delta L$. \Cref{fig:energy and angular momentum} shows that
this situation prevails in our simulation: $\Delta L<0$ and $\Delta\bar e>0$
generally for inner orbits, whereas $\Delta L>0$ and $\Delta\bar e<0$ generally
for outer orbits.

In fact, the rapid evolution of eccentricity distribution in
\cref{fig:eccentricity comparison}, exhibited by magnetized but not
unmagnetized eccentric disks, raises the question of whether any steady-state
structure exists for magnetized eccentric disks. Given the qualitative
similarity between stratified and unstratified disks, as well as those with
adiabatic and isothermal equations of state, it seems unlikely that disks with
properties different from ours would behave in qualitatively different ways
over the timescale we studied. In all the magnetized eccentric disks we
examined, the end result is a structure in which eccentricity declines outward,
but contains progressively less and less highly eccentric material as the most
eccentric material accretes rapidly onto the central object; such an effect is
manifest in \cref{fig:eccentricity comparison}. Gradual circularization spells
the end of a magnetized eccentric disk.

Strictly speaking, this argument applies to our simulated case of an isolated
disk and may be altered in the presence of external influences. Mass resupply
bringing in material with different energy and angular momentum could interfere
with eccentricity sorting. Gravitational perturbations from another object
could modify the energy and angular momentum budget of the disk. However,
because \ac{MHD}-driven evolution takes place on comparatively short
timescales, strong external influences would be needed to sustain a magnetized
eccentric disk in a steady state.

\subsection{Implications for \acsp*{TDE}}

In \acp{TDE}, the tidal gravity of supermassive black holes destroys stars
swooping by too close; for main-sequence stars, the critical distance is
$\mathrelp\approx27\,(GM/c^2)\times[M/(\qty{e6}{\solarmass})]^{-2/3}$, where
$M$ is the black hole mass \citep{2020ApJ...904...98R}. The bound stellar
debris is initially placed on very eccentric orbits with characteristic orbital
period $t_0\sim\qty{40}{\day}\times[M/(\qty{e6}{\solarmass})]^{1/2}$. Global
hydrodynamic simulations \citep{2015ApJ...804...85S, 2023ApJ...957...12R,
2024Natur.625..463S} revealed that over the first few $t_0$, shocks within the
debris transform the shape of the accretion flow into an irregular elliptical
disk with a mean eccentricity of \num{\sim0.5}, close to the eccentricity of
our simulated disk. The mean semimajor axis of this moderately eccentric disk
is only somewhat smaller than that of the debris shortly after disruption,
meaning that its orbital period is still $\mathrelp\sim t_0$. During the few
months when the disk is forming, the shocks can power optical/\ac{UV} flares
with luminosities and timescales roughly matching observed \ac{TDE} candidates
\citep{2015ApJ...806..164P}.

The initial condition of our simulation can be taken as representing, in
simplified form, the state of the irregular debris disk a few $t_0$ after
disruption. Our new simulation reveals two mechanisms due to vertical gravity
that can affect the \ac{TDE} light curve on timescales of several to ten $t_0$
after disk formation. Rapid dissipation of magnetic field near pericenter due
to vertical compression could power additional emission that may flatten the
decline of the optical/\ac{UV} flare of the \ac{TDE}, if the emission is in
that band. In addition, radial expansion of the disk, accelerated in the
presence of vertical gravity, could drive a small but potentially interesting
fraction of the debris to considerably smaller pericenters and higher
eccentricities.

The physics of the \ac{MRI} may also have observational implications beyond
$\mathrelp\sim10\,t_0$. The \ac{MRI} grows to nonlinear saturation in the
debris disk in \numrange{\sim5}{10} orbits with or without vertical gravity.
Consequently, \ac{MHD} stresses grow too slowly to drive much accretion onto
the black hole until after the main optical/\ac{UV} flare has faded. Instead,
they could power yearslong low-level emission, releasing at least as much
energy over time as the initial flare because accretion onto the black hole
should be at least as radiatively efficient as internal debris shocks. This
long-term accretion may explain the late-time emission
\citep[e.g.,][]{2019ApJ...878...82V, 2020ApJ...889..166J} or rebrightening
\citep[e.g.,][]{2022ApJ...928...63C, 2023A&A...669A..75L, 2023MNRAS.520.3549M,
2023ApJ...942L..33W} in a number of \ac{TDE} candidates.

\section{Conclusions}
\label{sec:conclusions}

Our earlier work established that the \ac{MRI} operates in unstratified
eccentric disks as vigorously as in circular disks; here we demonstrate that it
does in the presence of vertical gravity as well. Its nonlinear development in
a global simulation is qualitatively similar to circular disks, differing only
in particulars.

The mean Maxwell stress to pressure ratio $\alpha_{\su m}$ is similar to that
in circular disks at the order-of-magnitude level: The mean radial flux of
azimuthal momentum is outward and has magnitude $\num{\sim e-2}$ times the
pressure. Oscillation of the radial velocity between positive and negative
causes the Maxwell stress in a small region of the disk to transport angular
momentum inward rather than outward.

The strong orbital modulation of vertical gravity creates a brand new channel
for reconnection and dissipation, a direct consequence of the strong
compression near pericenter. Partly because of this, the magnetic energy is
weakened by a factor of \num{\sim3} over the course of our simulation.

\Ac{MHD} turbulence in eccentric disks drives rapid evolution of the radial
eccentricity profile, faster than purely hydrodynamic, secular processes. Over
only \numrange{\sim5}{10} orbits, stresses associated with \ac{MHD} turbulence
transport enough angular momentum to substantially widen the mass distribution
over eccentricity and create a strong radial eccentricity gradient with highly
eccentric material on the inside.

These \ac{MHD} effects can strongly influence eccentric disk structure on
timescales from a few to many orbits. During the several orbits required to
reach nonlinear saturation, they can reshuffle the eccentricity distribution;
once the nonlinear stage has been reached, the Maxwell stress created by
\ac{MRI}-driven turbulence can support accretion over longer periods. In
particular, in the eccentric disks formed by tidally disrupted stellar debris,
the \ac{MRI} may reach nonlinear saturation in merely \numrange{\sim5}{10}
orbits, but this can be long compared to the fallback time; therefore,
\ac{MRI}-driven effects may govern the accretion rate and luminosity on
timescales tens of times as long as the initial flare, accounting for the
emission years later.

\begin{acknowledgments}
The authors thank the anonymous referee for an astute question about the
influence of the equation of state. This material is based upon work supported
by the National Science Foundation (NSF) under grant AST-1908042 (CHC) and upon
work partially supported by the NSF under grants AST-2009260 and PHY-2110339
(JHK). TP was supported by the European Research Council Advanced Grant
\textquote{MultiJets} and a grant MP-SCMPS-00001470 from the Simons Foundation
to the Simons Collaboration on Extreme Electrodynamics of Compact Sources
(SCEECS). The simulation made use of the Extreme Science and Engineering
Discovery Environment (XSEDE) Stampede2 at the Texas Advanced Computing Center
(TACC) through allocation PHY210115; and the Advanced Research Computing at
Hopkins (ARCH) core facility (\url{https://www.arch.jhu.edu/}), which is
supported by the National Science Foundation (NSF) grant number OAC 1920103.
\end{acknowledgments}

\software{Athena++ \citep{2016ApJS..225...22W, 2020ApJS..249....4S}, NumPy
\citep{2020Natur.585..357H}, SymPy \citep{10.7717/peerj-cs.103}, Matplotlib
\citep{2007CSE.....9...90H}}

\begin{appendices}

\section{Metric components and Christoffel symbols}
\label{sec:metric}

The nonzero metric components in orbital coordinates are%
\allowdisplaybreaks
\begin{align}
g_{tt} &=
  -\symsf P, \\
g_{\lambda\lambda} &=
  r^2, \\
g_{\lambda\phi}=g_{\phi\lambda} &=
  R^2\symsf Q+z^2\symsf X, \\
g_{\lambda\zeta}=g_{\zeta\lambda} &=
  z^2/\zeta, \\
g_{\phi\phi} &=
  R^2(1+\symsf Q^2)+z^2\symsf X^2, \\
g_{\phi\zeta}=g_{\zeta\phi} &=
  z^2\symsf X/\zeta, \\
g_{\zeta\zeta} &=
  z^2/\zeta^2, \\
g^{tt} &=
  -1/\symsf P, \\
g^{\lambda\lambda} &=
  (1+\symsf Q^2)/R^2, \\
g^{\lambda\phi}=g^{\phi\lambda} &=
  -\symsf Q/R^2, \\
g^{\lambda\zeta}=g^{\zeta\lambda} &=
  -\zeta[1+\symsf Q(\symsf Q-\symsf X)]/R^2, \\
g^{\phi\phi} &=
  1/R^2, \\
g^{\phi\zeta}=g^{\zeta\phi} &=
  \zeta(\symsf Q-\symsf X)/R^2, \\
g^{\zeta\zeta} &=
  \zeta^2[r^2+z^2(\symsf Q-\symsf X)^2]/(R^2z^2),
\end{align}
the metric determinant is
\begin{equation}
g=-R^4z^2\symsf P/\zeta^2,
\end{equation}
and the nonzero Christoffel symbols of the second kind are
\begin{align}
\Gamma^t_{t\lambda}=\Gamma^t_{\lambda t}
  &= [(\partial_R\Phi)R+(\partial_z\Phi)z]/\symsf P, \\
\Gamma^t_{t\phi}=\Gamma^t_{\phi t}
  &= [(\partial_R\Phi)R\symsf Q+(\partial_z\Phi)z\symsf X]/\symsf P, \\
\Gamma^t_{t\zeta}=\Gamma^t_{\zeta t}
  &= (\partial_z\Phi)z/(\zeta\symsf P), \\
\Gamma^\lambda_{tt}
  &= \partial_R\Phi/R, \\
\Gamma^\lambda_{\lambda\lambda}
  &= 1, \\
\Gamma^\lambda_{\phi\phi}
  &= -R/\lambda, \\
\Gamma^\phi_{\lambda\phi}=\Gamma^\phi_{\phi\lambda}
  &= 1, \\
\Gamma^\phi_{\phi\phi}
  &= 2\symsf Q, \\
\Gamma^\zeta_{tt}
  &= \zeta(\partial_z\Phi/z-\partial_R\Phi/R), \\
\Gamma^\zeta_{\lambda\zeta}=\Gamma^\zeta_{\zeta\lambda}
  &= 1, \\
\Gamma^\zeta_{\phi\phi}
  &= \zeta[R/\lambda-(2\symsf Q-\symsf X)\symsf X+\symsf Y], \\
\Gamma^\zeta_{\phi\zeta}=\Gamma^\zeta_{\zeta\phi}
  &= \symsf X,
\end{align}
where
\begin{align}
\Phi(R,z) &\eqdef -1/(r+2), \\
r &\eqdef (R^2+z^2)^{1/2}, \\
\symsf P &\eqdef 1+2\Phi(R,z), \\
\symsf Q &\eqdef e\sin\phi/(1+e\cos\phi), \\
\symsf X &\eqdef \ods{\ln H_*}\phi, \\
\symsf Y &\eqdef \odds{\ln H_*}\phi.
\end{align}
\allowdisplaybreaks[0]

\section{Grid height}
\label{sec:grid height}

The grid height $H_*(\phi)$ controls how the grid stretches in the vertical
direction. As mentioned in \cref{sec:coordinate system}, we set $H_*(\phi)$ to
the scale height of a hydrostatic reference column positioned at some point
$(\lambda_*,\phi)$ along the characteristic orbit. The goal of this section is
to obtain an approximate closed-form expression for $H_*$; an exact measurement
of the scale height is not needed because $H_*$ does not affect disk physics.

We expand vertical gravity along $\lambda=\lambda_*$ to first order in $z$ as
\begin{equation}\label{eq:vertical frequency}
\evalb{-\partial_z\Phi}_{\lambda=\lambda_*}\approx-z/[R_*(R_*+2)^2]=-\nu_*^2z;
\end{equation}
here $\nu_*$ is the vertical frequency and $R_*\eqdef\lambda_*/(1+e\cos\phi)$.
The vertical density profile of the column is then
\citep[e.g.,][]{2021ApJ...920..130R}
\begin{equation}\label{eq:column density}
\rho_*(\phi,\zeta)\approx\rho_*(\phi,0)[1-(\gamma-1)z^2/H_*^2]^{1/(\gamma-1)},
\end{equation}
with the scale height of the column given by
\begin{equation}\label{eq:grid height 1}
H_*^2=2\gamma K_*\rho_*^{\gamma-1}(\phi,0)/\nu_*^2.
\end{equation}
\Cref{eq:column density,eq:grid height 1} reduce to their expected forms in the
isothermal limit of $\gamma\to1$.

Once we have picked a value for the surface density $\Sigma_*$ of the column,
we can solve for $\rho_*(\phi,0)$ and, through \cref{eq:grid height 1}, $H_*$.
In defining the surface density of the column, we break from the common
practice of integrating from the midplane to infinity, opting instead for
\begin{equation}
\Sigma_*=\int_0^{(\gamma-1)^{-1/2}H_*}dz\,\rho_*(\phi,\zeta),
\end{equation}
which has the advantage of smoothly connecting to the isothermal surface
density when $\gamma\to1$. Performing the integral and using \cref{eq:grid
height 1} to eliminate $\rho_*(\phi,0)$, we get
\begin{multline}\label{eq:grid height 2}
\Sigma_*=(\gamma-1)^{-1/2}(2\gamma K_*)^{-1/(\gamma-1)}
  \mathop{_2F_1}(\tfrac12,-(\gamma-1)^{-1},\tfrac32;1)\times{} \\
H_*^{(\gamma+1)/(\gamma-1)}\nu_*^{2/(\gamma-1)},
\end{multline}
where $\mathop{_2F_1}$ is the hypergeometric function.

The metric and connection depend on the derivatives of $H_*$ as well.
Differentiating \cref{eq:grid height 2} furnishes us with
\begin{equation}
\frac{d^n}{d\phi^n}\ln H_*=-\frac2{\gamma+1}\frac{d^n}{d\phi^n}\ln\nu_*,
\end{equation}
where $n$ is a positive integer, and differentiating \cref{eq:vertical
frequency} yields
\begin{align}
\od{}\phi\ln\nu_* &=
  -\frac{3R_*+2}{2(R_*+2)}\symsf Q, \\
\odd{}\phi\ln\nu_* &=
  -\frac{3R_*+2}{2(R_*+2)}\od{\symsf Q}\phi-\frac{2R_*}{(R_*+2)^2}\symsf Q^2
\end{align}
with $\symsf Q$ from \cref{sec:metric}.

\section{Initial eccentric disk}
\label{sec:deformation}

The construction of an eccentric disk in approximate force balance begins with
an exactly hydrostatic circular torus. This reference torus is described in
cylindrical coordinates $(R,\varphi,z)$ and is governed by five parameters
$(\tilde R,\tilde K,\tilde\Gamma,\tilde\rho_{\su m},\tilde q)$. Its density
$\tilde\rho(R,z)$ and pressure $\tilde p(R,z)$ are related by the polytropic
equation of state $\tilde p=\tilde K\tilde\rho^{\tilde\Gamma}$, and its density
maximum is $\tilde\rho_{\su m}=\tilde\rho(\tilde R,0)$. Its orbital velocity
follows the shear profile $\tilde{\vec v}(R)=\tilde v_{\su m}(R/\tilde
R)^{1-\tilde q}\,\uvec e_\varphi$, where the azimuthal velocity at the density
maximum is given by $\tilde v_{\su m}^2=(R\partial_R\Phi)(\tilde R,0)$. With
these stipulations, $\tilde\rho$ is the solution to the equation \citep[see
also][]{1984MNRAS.208..721P, 2000ApJ...528..462H}
\begin{equation}
\const=\Phi-\frac{\tilde v_\varphi^2}{2(1-\tilde q)}
+\begin{cases}
  \tilde K\tfrac{\tilde\Gamma}{\tilde\Gamma-1}(\tilde\rho^{\tilde\Gamma-1}-1),
    & \tilde\Gamma\ne1, \\
  \tilde K\ln\tilde\rho, & \tilde\Gamma=1.
\end{cases}
\end{equation}
We demand $\tilde R=\lambda_*$ so that the torus matches the desired initial
eccentric disk in characteristic semilatus rectum, and $(\tilde
K,\tilde\Gamma,\tilde q)=(\num{9e-6},1,1.55)$ so that the torus is
geometrically thin. We also arbitrarily set $\tilde\rho_{\su m}=1$. The surface
density of such a torus at $R=\tilde R$ is
\begin{equation}
\tilde\Sigma=\int_0dz\,\tilde\rho(\tilde R,z)\approx8.30,
\end{equation}
the upper limit being the height at which the integrand vanishes.

The circular Newtonian torus tells us how to set up in our simulation a
circular \ac{GR} disk in the weak-gravity limit. We employ eccentric
coordinates with $e=0$ for the initial condition in anticipation of the
subsequent conversion of the circular disk to an eccentric one. The vertical
scaling of the eccentric coordinate system is governed by a reference column
with $(\Sigma_*,K_*)=(\tilde\Sigma,\tilde K)$, which implies
$H_*(\phi)\approx11.9$ for all $\phi$. \Cref{eq:transform 1,eq:transform
2,eq:transform 3} are used to transform the torus from the cylindrical
coordinates in which it is described to eccentric coordinates.

Torus properties must be converted to primitive variables for consumption by
the simulation. Density and pressure are directly usable as primitive
variables. The Newtonian angular velocity $\tilde v_\varphi/R$ is identified
with the \ac{GR} coordinate velocity $u^\phi/u^t$; together with the constraint
$u^\lambda=u^\zeta=0$ and the normalization condition, the velocity is fully
determined.

Once the circular disk is configured on the eccentric coordinate system, it is
morphed into an eccentric disk by increasing the eccentricity $e$ of the
coordinate system itself. Each cell stays at its coordinate-space location
$(\lambda,\phi,\zeta)$ but shifts to a new physical-space location
$(R,\varphi,z)$. We use the same reference column as above to define the
breathing of the new grid, but now $H_*(\phi)$ is a function of $\phi$ because
different points along the characteristic orbit are at different $R_*$ and
experience different levels of vertical gravity. Since we do not touch the grid
in coordinate space, we need not interpolate simulation data. We also get
vertical motion for free: Grid geometry forces the velocity field to diverge
and converge in synchronization with disk expansion and contraction even though
$u^\lambda=u^\zeta=0$.

Two additional tweaks keep the eccentric disk as hydrostatic as possible. Disk
reshaping moves material to a different $R$, breaking the balance between
gravitational and centrifugal forces. Restoring that balance involves
multiplying $u^\phi$ of each cell by $R^2/\lambda^2$ so that material has the
same specific angular momentum as before. Reshaping also changes the disk
height, so the vertically integrated mass current around an orbit of constant
$\lambda$ is no longer constant over $\phi$. To approximately reinstate mass
continuity, we scale $\rho$ in any given column of cells by the same factor so
that the surface density of the column equals its value before reshaping. We
multiply $p$ by that factor as well.

An eccentric disk assembled with the aforementioned process is not strictly
hydrostatic, but close enough for the study of the \ac{MRI}. Instead of jumping
immediately to the desired eccentricity, one could raise eccentricity stepwise,
each step of reshaping followed by a period of relaxation, but we found that
this does not yield a quieter disk.

\section{Instantaneous eccentricity}
\label{sec:eccentricity}

In \citet{2022ApJ...933...81C}, the instantaneous eccentricity $\bar e$ of a
gas packet in an unstratified disk is defined as the eccentricity of a particle
orbit having the same specific energy $E$ and specific angular momentum $L$ as
the gas packet. That definition does not work for a stratified disk because the
trajectory of a gas packet does not in general lie in some orbital plane that
includes the central object, not even if we smooth out turbulent motion by
considering a time-averaged trajectory. This necessitates an alternative
approach.

We consider a gas packet at some $(t,\lambda,\phi,\zeta)$. The velocity of the
gas packet has physical, cylindrical components $u^{\hat i}$, obtained from its
contravariant components $u^i$ using equations analogous to
\cref{eq:cylindrical magnetic field 1,eq:cylindrical magnetic field
2,eq:cylindrical magnetic field 3}. We construct the restriction of this
velocity to the midplane, by which we mean a properly normalized velocity in
the tangent space of $(t,\lambda,\phi,0)$ that has the same coordinate velocity
as $u^\mu$ except in the $\zeta$\nobreakdash-direction. Let us denote
quantities at $(t,\lambda,\phi,0)$ by a breve. The construction guarantees a
unique solution for the contravariant temporal component $\breve u^t$ and the
physical, cylindrical components $\breve u^{\hat i}$ of the restricted
velocity:
\begin{align}
\breve u^{\hat R}/\breve u^t &= u^{\hat R}/u^t, \\
\breve u^{\hat\varphi}/\breve u^t &= u^{\hat\varphi}/u^t, \\
\breve u^{\hat z} &= 0, \\
\breve g_{tt}\breve u^t\breve u^t+\breve u^{\hat R}\breve u^{\hat R}
  +\breve u^{\hat\varphi}\breve u^{\hat\varphi}
  +\breve u^{\hat z}\breve u^{\hat z} &= -1.
\end{align}
The specific energy and angular momentum of the restricted velocity are
$E=\breve g_{tt}\breve u^t$ and $L=R\breve u^{\hat\varphi}$, respectively;
these are the quantities plotted in \cref{fig:energy and angular momentum}. The
instantaneous eccentricity $\bar e$ is then obtained in the same way as in
\citet{2022ApJ...933...81C}:
\begin{equation}
\bar e^2=1+(E^2-1)L^2/E^4.
\end{equation}

\end{appendices}

\ifapj\bibliography{mri}\fi

@article{10.7717/peerj-cs.103,
  author        = "Meurer, Aaron and Smith, Christopher P. and Paprocki, Mateusz and \v{C}ert\'{i}k, Ond\v{r}ej and Kirpichev, Sergey B. and Rocklin, Matthew and Kumar, AMiT and Ivanov, Sergiu and Moore, Jason K. and Singh, Sartaj and Rathnayake, Thilina and Vig, Sean and Granger, Brian E. and Muller, Richard P. and Bonazzi, Francesco and Gupta, Harsh and Vats, Shivam and Johansson, Fredrik and Pedregosa, Fabian and Curry, Matthew J. and Terrel, Andy R. and Rou\v{c}ka, \v{S}t\v{e}p\'{a}n and Saboo, Ashutosh and Fernando, Isuru and Kulal, Sumith and Cimrman, Robert and Scopatz, Anthony",
  title         = "SymPy: symbolic computing in Python",
  year          = 2017,
  month         = 1,
  journal       = "PeerJ Comput. Sci.",
  volume        = 585,
  pages         = "e103",
  doi           = "10.7717/peerj-cs.103"
}

@article{1973A&A....24..337S,
  author        = "Shakura, N.~I. and Sunyaev, R.~A.",
  title         = "Black holes in binary systems. Observational appearance",
  year          = 1973,
  month         = 6,
  journal       = "A\&A",
  volume        = 24,
  pages         = "337--355",
  bibcode       = "1973A&A....24..337S"
}

@article{1978MNRAS.185..629K,
  author        = "Kato, Shoji",
  title         = "Pulsational instability of accretion disks to axially symmetric oscillations",
  year          = 1978,
  month         = 12,
  journal       = "MNRAS",
  volume        = 185,
  pages         = "629--642",
  doi           = "10.1093/mnras/185.3.629",
  bibcode       = "1978MNRAS.185..629K"
}

@article{1984MNRAS.208..721P,
  author        = "Papaloizou, J.~C.~B. and Pringle, J.~E.",
  title         = "The dynamical stability of differentially rotating discs with constant specific angular momentum",
  year          = 1984,
  month         = 6,
  journal       = "MNRAS",
  volume        = 208,
  pages         = "721--750",
  doi           = "10.1093/mnras/208.4.721",
  bibcode       = "1984MNRAS.208..721P"
}

@article{1988MNRAS.232...35W,
  author        = "Whitehurst, Robert",
  title         = "Numerical simulations of accretion discs - I. Superhumps : a tidal phenomenon of accretion discs.",
  year          = 1988,
  month         = 5,
  journal       = "MNRAS",
  volume        = 232,
  pages         = "35--51",
  doi           = "10.1093/mnras/232.1.35",
  bibcode       = "1988MNRAS.232...35W"
}

@article{1991ApJ...376..214B,
  author        = "Balbus, Steven A. and Hawley, John F.",
  title         = "A Powerful Local Shear Instability in Weakly Magnetized Disks. I. Linear Analysis",
  year          = 1991,
  month         = 7,
  journal       = "ApJ",
  volume        = 376,
  pages         = "214--222",
  doi           = "10.1086/170270",
  bibcode       = "1991ApJ...376..214B"
}

@article{1991ApJ...376..223H,
  author        = "Hawley, John F. and Balbus, Steven A.",
  title         = "A Powerful Local Shear Instability in Weakly Magnetized Disks. II. Nonlinear Evolution",
  year          = 1991,
  month         = 7,
  journal       = "ApJ",
  volume        = 376,
  pages         = "223--233",
  doi           = "10.1086/170271",
  bibcode       = "1991ApJ...376..223H"
}

@article{1991ApJ...381..259L,
  author        = "Lubow, Stephen H.",
  title         = "A Model for Tidally Driven Eccentric Instabilities in Fluid Disks",
  year          = 1991,
  month         = 11,
  journal       = "ApJ",
  volume        = 381,
  pages         = "259--267",
  doi           = "10.1086/170647",
  bibcode       = "1991ApJ...381..259L"
}

@article{1994ApJ...432..213G,
  author        = "Goodman, Jeremy and Xu, Guohong",
  title         = "Parasitic Instabilities in Magnetized, Differentially Rotating Disks",
  year          = 1994,
  month         = 9,
  journal       = "ApJ",
  volume        = 432,
  pages         = "213--223",
  doi           = "10.1086/174562",
  bibcode       = "1994ApJ...432..213G"
}

@article{1994MNRAS.266..583L,
  author        = "Lyubarskij, Y.~E. and Postnov, K.~A. and Prokhorov, M.~E.",
  title         = "Eccentric Accretion Discs",
  year          = 1994,
  month         = 2,
  journal       = "MNRAS",
  volume        = 266,
  pages         = "583--596",
  doi           = "10.1093/mnras/266.3.583",
  bibcode       = "1994MNRAS.266..583L"
}

@article{1995ApJ...438..610E,
  author        = "Eracleous, Michael and Livio, Mario and Halpern, Jules P. and Storchi-Bergmann, Thaisa",
  title         = "Elliptical Accretion Disks in Active Galactic Nuclei",
  year          = 1995,
  month         = 1,
  journal       = "ApJ",
  volume        = 438,
  pages         = "610--622",
  doi           = "10.1086/175104",
  bibcode       = "1995ApJ...438..610E"
}

@article{1995ApJ...440..742H,
  author        = "Hawley, John F. and Gammie, Charles F. and Balbus, Steven A.",
  title         = "Local Three-dimensional Magnetohydrodynamic Simulations of Accretion Disks",
  year          = 1995,
  month         = 2,
  journal       = "ApJ",
  volume        = 440,
  pages         = "742--763",
  doi           = "10.1086/175311",
  bibcode       = "1995ApJ...440..742H"
}

@article{1998RvMP...70....1B,
  author        = "Balbus, Steven A. and Hawley, John F.",
  title         = "Instability, turbulence, and enhanced transport in accretion disks",
  year          = 1998,
  month         = 1,
  journal       = "RvMP",
  volume        = 70,
  pages         = "1--53",
  doi           = "10.1103/RevModPhys.70.1",
  bibcode       = "1998RvMP...70....1B"
}

@article{1999ApJ...517..700L,
  author        = "Lazarian, A. and Vishniac, Ethan T.",
  title         = "Reconnection in a Weakly Stochastic Field",
  year          = 1999,
  month         = 6,
  journal       = "ApJ",
  volume        = 517,
  pages         = "700--718",
  doi           = "10.1086/307233",
  archivePrefix = "arXiv",
  eprint        = "astro-ph/9811037",
  bibcode       = "1999ApJ...517..700L"
}

@article{2000ApJ...528..462H,
  author        = "Hawley, John F.",
  title         = "Global Magnetohydrodynamical Simulations of Accretion Tori",
  year          = 2000,
  month         = 1,
  journal       = "ApJ",
  volume        = 528,
  pages         = "462--479",
  doi           = "10.1086/308180",
  archivePrefix = "arXiv",
  eprint        = "astro-ph/9907385",
  bibcode       = "2000ApJ...528..462H"
}

@book{2000ssd..book.....M,
  author        = "Murray, Carl D. and Dermott, Stanley F.",
  title         = "Solar System Dynamics",
  year          = 1999,
  address       = "Cambridge",
  publisher     = "Cambridge University Press",
  bibcode       = "2000ssd..book.....M"
}

@article{2001AJ....122.2257S,
  author        = "Statler, Thomas S.",
  title         = "A Simple Family of Models for Eccentric Keplerian Fluid Disks",
  year          = 2001,
  month         = 11,
  journal       = "AJ",
  volume        = 122,
  pages         = "2257--2266",
  doi           = "10.1086/323713",
  archivePrefix = "arXiv",
  eprint        = "astro-ph/0108095",
  bibcode       = "2001AJ....122.2257S"
}

@article{2001MNRAS.325..231O,
  author        = "Ogilvie, G.~I.",
  title         = "Non-linear fluid dynamics of eccentric discs",
  year          = 2001,
  month         = 7,
  journal       = "MNRAS",
  volume        = 325,
  pages         = "231--248",
  doi           = "10.1046/j.1365-8711.2001.04416.x",
  archivePrefix = "arXiv",
  eprint        = "astro-ph/0102245",
  bibcode       = "2001MNRAS.325..231O"
}

@article{2002ApJ...566..164H,
  author        = "Hawley, John F. and Krolik, Julian H.",
  title         = "High-Resolution Simulations of the Plunging Region in a Pseudo-Newtonian Potential: Dependence on Numerical Resolution and Field Topology",
  year          = 2002,
  month         = 2,
  journal       = "ApJ",
  volume        = 566,
  pages         = "164--180",
  doi           = "10.1086/338059",
  archivePrefix = "arXiv",
  eprint        = "astro-ph/0110118",
  bibcode       = "2002ApJ...566..164H"
}

@article{2003ApJ...589..444G,
  author        = "Gammie, Charles F. and McKinney, Jonathan C. and T{\'o}th, G{\'a}bor",
  title         = "HARM: A Numerical Scheme for General Relativistic Magnetohydrodynamics",
  year          = 2003,
  month         = 5,
  journal       = "ApJ",
  volume        = 589,
  pages         = "444--457",
  doi           = "10.1086/374594",
  archivePrefix = "arXiv",
  eprint        = "astro-ph/0301509",
  bibcode       = "2003ApJ...589..444G"
}

@article{2005A&A...432..743P,
  author        = "Papaloizou, J.~C.~B.",
  title         = "The local instability of steady astrophysical flows with non circular streamlines with application to differentially rotating disks with free eccentricity",
  year          = 2005,
  month         = 3,
  journal       = "A{\&}A",
  volume        = 432,
  pages         = "743--755",
  doi           = "10.1051/0004-6361:20041947",
  archivePrefix = "arXiv",
  eprint        = "astro-ph/0412587",
  bibcode       = "2005A&A...432..743P"
}

@article{2005A&A...432..757P,
  author        = "Papaloizou, J.~C.~B.",
  title         = "Global numerical simulations of differentially rotating disks with free eccentricity",
  year          = 2005,
  month         = 3,
  journal       = "A{\&}A",
  volume        = 432,
  pages         = "757--769",
  doi           = "10.1051/0004-6361:20041948",
  bibcode       = "2005A&A...432..757P"
}

@article{2006ApJ...640..901H,
  author        = "Hirose, Shigenobu and Krolik, Julian H. and Stone, James M.",
  title         = "Vertical Structure of Gas Pressure-dominated Accretion Disks with Local Dissipation of Turbulence and Radiative Transport",
  year          = 2006,
  month         = 4,
  journal       = "ApJ",
  volume        = 640,
  pages         = "901--917",
  doi           = "10.1086/499153",
  archivePrefix = "arXiv",
  eprint        = "astro-ph/0510741",
  bibcode       = "2006ApJ...640..901H"
}

@article{2006Sci...314.1908G,
  author        = "G{\"a}nsicke, B.~T. and Marsh, T.~R. and Southworth, J. and Rebassa-Mansergas, A.",
  title         = "A Gaseous Metal Disk Around a White Dwarf",
  year          = 2006,
  month         = 12,
  journal       = "Sci",
  volume        = 314,
  pages         = "1908--1910",
  doi           = "10.1126/science.1135033",
  archivePrefix = "arXiv",
  eprint        = "astro-ph/0612697",
  bibcode       = "2006Sci...314.1908G"
}

@article{2007CSE.....9...90H,
  author        = "Hunter, John D.",
  title         = "Matplotlib: A 2D Graphics Environment",
  year          = 2007,
  month         = 5,
  journal       = "CSE",
  volume        = 9,
  pages         = "90--95",
  doi           = "10.1109/MCSE.2007.55",
  bibcode       = "2007CSE.....9...90H"
}

@article{2008MNRAS.388.1372O,
  author        = "Ogilvie, Gordon I.",
  title         = "3D eccentric discs around Be stars",
  year          = 2008,
  month         = 8,
  journal       = "MNRAS",
  volume        = 388,
  pages         = "1372--1380",
  doi           = "10.1111/j.1365-2966.2008.13484.x",
  archivePrefix = "arXiv",
  eprint        = "0805.3128",
  bibcode       = "2008MNRAS.388.1372O"
}

@article{2008Sci...321.1060B,
  author        = "Bonnell, I.~A. and Rice, W.~K.~M.",
  title         = "Star Formation Around Supermassive Black Holes",
  year          = 2008,
  month         = 8,
  journal       = "Sci",
  volume        = 321,
  pages         = "1060--1062",
  doi           = "10.1126/science.1160653",
  archivePrefix = "arXiv",
  eprint        = "0810.2723",
  bibcode       = "2008Sci...321.1060B"
}

@article{2011ApJ...738...84H,
  author        = "Hawley, John F. and Guan, Xiaoyue and Krolik, Julian H.",
  title         = "Assessing Quantitative Results in Accretion Simulations: From Local to Global",
  year          = 2011,
  month         = 9,
  journal       = "ApJ",
  volume        = 738,
  pages         = 84,
  doi           = "10.1088/0004-637X/738/1/84",
  archivePrefix = "arXiv",
  eprint        = "1103.5987",
  bibcode       = "2011ApJ...738...84H"
}

@article{2013ApJ...772..102H,
  author        = "Hawley, John F. and Richers, Sherwood A. and Guan, Xiaoyue and Krolik, Julian H.",
  title         = "Testing Convergence for Global Accretion Disks",
  year          = 2013,
  month         = 8,
  journal       = "ApJ",
  volume        = 772,
  pages         = 102,
  doi           = "10.1088/0004-637X/772/2/102",
  archivePrefix = "arXiv",
  eprint        = "1306.0243",
  bibcode       = "2013ApJ...772..102H"
}

@article{2014ApJ...783...23G,
  author        = "Guillochon, James and Manukian, Haik and Ramirez-Ruiz, Enrico",
  title         = "PS1-10jh: The Disruption of a Main-sequence Star of Near-solar Composition",
  year          = 2014,
  month         = 3,
  journal       = "ApJ",
  volume        = 783,
  pages         = 23,
  doi           = "10.1088/0004-637X/783/1/23",
  archivePrefix = "arXiv",
  eprint        = "1304.6397",
  bibcode       = "2014ApJ...783...23G"
}

@article{2014MNRAS.445.2621O,
  author        = "Ogilvie, Gordon I. and Barker, Adrian J.",
  title         = "Local and global dynamics of eccentric astrophysical discs",
  year          = 2014,
  month         = 12,
  journal       = "MNRAS",
  volume        = 445,
  pages         = "2621--2636",
  doi           = "10.1093/mnras/stu1795",
  archivePrefix = "arXiv",
  eprint        = "1409.6487",
  bibcode       = "2014MNRAS.445.2621O"
}

@article{2015ApJ...804...85S,
  author        = "Shiokawa, Hotaka and Krolik, Julian H. and Cheng, Roseanne M. and Piran, Tsvi and Noble, Scott C.",
  title         = "General Relativistic Hydrodynamic Simulation of Accretion Flow from a Stellar Tidal Disruption",
  year          = 2015,
  month         = 5,
  journal       = "ApJ",
  volume        = 804,
  pages         = 85,
  doi           = "10.1088/0004-637X/804/2/85",
  archivePrefix = "arXiv",
  eprint        = "1501.04365",
  bibcode       = "2015ApJ...804...85S"
}

@article{2015ApJ...806..164P,
  author        = "Piran, Tsvi and Svirski, Gilad and Krolik, Julian and Cheng, Roseanne M. and Shiokawa, Hotaka",
  title         = "‧Disk Formation Versus Disk Accretion—What Powers Tidal Disruption Events?",
  year          = 2015,
  month         = 6,
  journal       = "ApJ",
  volume        = 806,
  pages         = 164,
  doi           = "10.1088/0004-637X/806/2/164",
  archivePrefix = "arXiv",
  eprint        = "1502.05792",
  bibcode       = "2015ApJ...806..164P"
}

@article{2016ApJS..225...22W,
  author        = "White, Christopher J. and Stone, James M. and Gammie, Charles F.",
  title         = "An Extension of the Athena++ Code Framework for GRMHD Based on Advanced Riemann Solvers and Staggered-mesh Constrained Transport",
  year          = 2016,
  month         = 8,
  journal       = "ApJS",
  volume        = 225,
  pages         = 22,
  doi           = "10.3847/0067-0049/225/2/22",
  archivePrefix = "arXiv",
  eprint        = "1511.00943",
  bibcode       = "2016ApJS..225...22W"
}

@article{2016MNRAS.458.3221T,
  author        = "Teyssandier, Jean and Ogilvie, Gordon I.",
  title         = "Growth of eccentric modes in disc-planet interactions",
  year          = 2016,
  month         = 5,
  journal       = "MNRAS",
  volume        = 458,
  pages         = "3221--3247",
  doi           = "10.1093/mnras/stw521",
  archivePrefix = "arXiv",
  eprint        = "1603.00653",
  bibcode       = "2016MNRAS.458.3221T"
}

@article{2017MNRAS.467.1426S,
  author        = "Svirski, Gilad and Piran, Tsvi and Krolik, Julian",
  title         = "Elliptical Accretion and Low Luminosity from High Accretion Rate Stellar Tidal Disruption Events",
  year          = 2017,
  month         = 5,
  journal       = "MNRAS",
  volume        = 467,
  pages         = "1426--1432",
  doi           = "10.1093/mnras/stx117",
  archivePrefix = "arXiv",
  eprint        = "1508.02389",
  bibcode       = "2017MNRAS.467.1426S"
}

@article{2017MNRAS.472L..99L,
  author        = "Liu, F.~K. and Zhou, Z.~Q. and Cao, R. and Ho, L.~C. and Komossa, S.",
  title         = "Disc origin of broad optical emission lines of the TDE candidate PTF09djl",
  year          = 2017,
  month         = 11,
  journal       = "MNRAS",
  volume        = 472,
  pages         = "L99--L103",
  doi           = "10.1093/mnrasl/slx147",
  archivePrefix = "arXiv",
  eprint        = "1710.04229",
  bibcode       = "2017MNRAS.472L..99L"
}

@article{2018ApJ...856...12C,
  author        = "Chan, Chi-Ho and Krolik, Julian H. and Piran, Tsvi",
  title         = "Magnetorotational Instability in Eccentric Disks",
  year          = 2018,
  month         = 3,
  journal       = "ApJ",
  volume        = 856,
  pages         = 12,
  doi           = "10.3847/1538-4357/aab15c",
  archivePrefix = "arXiv",
  eprint        = "1712.01882",
  bibcode       = "2018ApJ...856...12C"
}

@article{2019ApJ...878...82V,
  author        = "van Velzen, Sjoert and Stone, Nicholas C. and Metzger, Brian D. and Gezari, Suvi and Brown, Thomas M. and Fruchter, Andrew S.",
  title         = "Late-time UV Observations of Tidal Disruption Flares Reveal Unobscured, Compact Accretion Disks",
  year          = 2019,
  month         = 6,
  journal       = "ApJ",
  volume        = 878,
  pages         = 82,
  doi           = "10.3847/1538-4357/ab1844",
  archivePrefix = "arXiv",
  eprint        = "1809.00003",
  bibcode       = "2019ApJ...878...82V"
}

@article{2020ApJ...889..166J,
  author        = "Jonker, P.~G. and Stone, N.~C. and Generozov, A. and van Velzen, S. and Metzger, B.",
  title         = "Implications from Late-time X-Ray Detections of Optically Selected Tidal Disruption Events: State Changes, Unification, and Detection Rates",
  year          = 2020,
  month         = 2,
  journal       = "ApJ",
  volume        = 889,
  pages         = 166,
  doi           = "10.3847/1538-4357/ab659c",
  archivePrefix = "arXiv",
  eprint        = "1906.12236",
  bibcode       = "2020ApJ...889..166J"
}

@article{2020ApJ...904...98R,
  author        = "Ryu, Taeho and Krolik, Julian and Piran, Tsvi and Noble, Scott C.",
  title         = "Tidal Disruptions of Main-sequence Stars. I. Observable Quantities and Their Dependence on Stellar and Black Hole Mass",
  year          = 2020,
  month         = 12,
  journal       = "ApJ",
  volume        = 904,
  pages         = 98,
  doi           = "10.3847/1538-4357/abb3cf",
  archivePrefix = "arXiv",
  eprint        = "2001.03501",
  bibcode       = "2020ApJ...904...98R"
}

@article{2020ApJS..249....4S,
  author        = "Stone, James M. and Tomida, Kengo and White, Christopher J. and Felker, Kyle G.",
  title         = "The Athena++ Adaptive Mesh Refinement Framework: Design and Magnetohydrodynamic Solvers",
  year          = 2020,
  month         = 7,
  journal       = "ApJS",
  volume        = 249,
  pages         = 4,
  doi           = "10.3847/1538-4365/ab929b",
  archivePrefix = "arXiv",
  eprint        = "2005.06651",
  bibcode       = "2020ApJS..249....4S"
}

@article{2020MNRAS.497..451D,
  author        = "Dewberry, Janosz W. and Latter, Henrik N. and Ogilvie, Gordon I. and Fromang, Sebastien",
  title         = "HFQPOs and discoseismic mode excitation in eccentric, relativistic discs. II. Magnetohydrodynamic simulations",
  year          = 2020,
  month         = 9,
  journal       = "MNRAS",
  volume        = 497,
  pages         = "451--465",
  doi           = "10.1093/mnras/staa1898",
  archivePrefix = "arXiv",
  eprint        = "2006.16266",
  bibcode       = "2020MNRAS.497..451D"
}

@article{2020Natur.585..357H,
  author        = "Harris, Charles R. and Millman, K.~Jarrod and van der Walt, St{\'e}fan J. and Gommers, Ralf and Virtanen, Pauli and Cournapeau, David and Wieser, Eric and Taylor, Julian and Berg, Sebastian and Smith, Nathaniel J. and Kern, Robert and Picus, Matti and Hoyer, Stephan and van Kerkwijk, Marten H. and Brett, Matthew and Haldane, Allan and del R{\'i}o, Jaime Fern{\'a}ndez and Wiebe, Mark and Peterson, Pearu and G{\'e}rard-Marchant, Pierre and Sheppard, Kevin and Reddy, Tyler and Weckesser, Warren and Abbasi, Hameer and Gohlke, Christoph and Oliphant, Travis E.",
  title         = "Array programming with NumPy",
  year          = 2020,
  month         = 9,
  journal       = "Natur",
  volume        = 585,
  pages         = "357--362",
  doi           = "10.1038/s41586-020-2649-2",
  archivePrefix = "arXiv",
  eprint        = "2006.10256",
  bibcode       = "2020Natur.585..357H"
}

@article{2021ApJ...920..130R,
  author        = "Ryu, Taeho and Krolik, Julian and Piran, Tsvi",
  title         = "The Impact of Shocks on the Vertical Structure of Eccentric Disks",
  year          = 2021,
  month         = 10,
  journal       = "ApJ",
  volume        = 920,
  pages         = 130,
  doi           = "10.3847/1538-4357/ac185a",
  archivePrefix = "arXiv",
  eprint        = "2105.09434",
  bibcode       = "2021ApJ...920..130R"
}

@article{2021MNRAS.500.4110L,
  author        = "Lynch, Elliot M. and Ogilvie, Gordon I.",
  title         = "Dynamical structure of highly eccentric discs with applications to tidal disruption events",
  year          = 2021,
  month         = 1,
  journal       = "MNRAS",
  volume        = 500,
  pages         = "4110--4125",
  doi           = "10.1093/mnras/staa3459",
  archivePrefix = "arXiv",
  eprint        = "2011.02219",
  bibcode       = "2021MNRAS.500.4110L"
}

@article{2021MNRAS.501.5500L,
  author        = "Lynch, Elliot M. and Ogilvie, Gordon I.",
  title         = "Importance of magnetic fields in highly eccentric discs with applications to tidal disruption events",
  year          = 2021,
  month         = 3,
  journal       = "MNRAS",
  volume        = 501,
  pages         = "5500--5516",
  doi           = "10.1093/mnras/staa4026",
  archivePrefix = "arXiv",
  eprint        = "2101.01221",
  bibcode       = "2021MNRAS.501.5500L"
}

@article{2021MNRAS.505L..21T,
  author        = "Trevascus, David and Price, Daniel J. and Nealon, Rebecca and Liptai, David and Manser, Christopher J. and Veras, Dimitri",
  title         = "Formation of eccentric gas discs from sublimating or partially disrupted asteroids orbiting white dwarfs",
  year          = 2021,
  month         = 7,
  journal       = "MNRAS",
  volume        = 505,
  pages         = "L21--L25",
  doi           = "10.1093/mnrasl/slab043",
  archivePrefix = "arXiv",
  eprint        = "2105.00626",
  bibcode       = "2021MNRAS.505L..21T"
}

@article{2021MNRAS.506.6014T,
  author        = "Tucker, M.~A. and Shappee, B.~J. and Hinkle, J.~T. and Neustadt, J.~M.~M. and Eracleous, M. and Kochanek, C.~S. and Prieto, J.~L. and Payne, A.~V. and Galbany, L. and Anderson, J.~P. and Auchettl, K. and Auge, C. and Holoien, Thomas W.-S.",
  title         = "An AMUSING look at the host of the periodic nuclear transient ASASSN-14ko reveals a second AGN",
  year          = 2021,
  month         = 10,
  journal       = "MNRAS",
  volume        = 506,
  pages         = "6014--6028",
  doi           = "10.1093/mnras/stab2085",
  archivePrefix = "arXiv",
  eprint        = "2011.05998",
  bibcode       = "2021MNRAS.506.6014T"
}

@article{2022A&A...659A..91W,
  author        = "Wissing, Robert and Shen, Sijing and Wadsley, James and Quinn, Thomas",
  title         = "Magnetorotational instability with smoothed particle hydrodynamics",
  year          = 2022,
  month         = 3,
  journal       = "A{\&}A",
  volume        = 659,
  pages         = "A91",
  doi           = "10.1051/0004-6361/202141206",
  archivePrefix = "arXiv",
  eprint        = "2105.01091",
  bibcode       = "2022A&A...659A..91W"
}

@article{2022ApJ...928...63C,
  author        = "Chen, Jin-Hong and Dou, Li-Ming and Shen, Rong-Feng",
  title         = "AT 2019avd: A Tidal Disruption Event with a Two-phase Evolution",
  year          = 2022,
  month         = 3,
  journal       = "ApJ",
  volume        = 928,
  pages         = 63,
  doi           = "10.3847/1538-4357/ac558d",
  archivePrefix = "arXiv",
  eprint        = "2106.08835",
  bibcode       = "2022ApJ...928...63C"
}

@article{2022ApJ...933...81C,
  author        = "Chan, Chi-Ho and Piran, Tsvi and Krolik, Julian H.",
  title         = "Nonlinear Evolution of the Magnetorotational Instability in Eccentric Disks",
  year          = 2022,
  month         = 7,
  journal       = "ApJ",
  volume        = 933,
  pages         = 81,
  doi           = "10.3847/1538-4357/ac68f3",
  archivePrefix = "arXiv",
  eprint        = "2201.03728",
  bibcode       = "2022ApJ...933...81C"
}

@article{2022MNRAS.517.2639Z,
  author        = "Zier, Oliver and Springel, Volker",
  title         = "Simulating the magnetorotational instability on a moving mesh with the shearing box approximation",
  year          = 2022,
  month         = 12,
  journal       = "MNRAS",
  volume        = 517,
  pages         = "2639--2658",
  doi           = "10.1093/mnras/stac2831",
  archivePrefix = "arXiv",
  eprint        = "2208.01065",
  bibcode       = "2022MNRAS.517.2639Z"
}

@article{2023A&A...669A..75L,
  author        = "Liu, Z. and Malyali, A. and Krumpe, M. and Homan, D. and Goodwin, A.~J. and Grotova, I. and Kawka, A. and Rau, A. and Merloni, A. and Anderson, G.~E. and Miller-Jones, J.~C.~A. and Markowitz, A.~G. and Ciroi, S. and Di Mille, F. and Schramm, M. and Tang, S. and Buckley, D.~A.~H. and Gromadzki, M. and Jin, C. and Buchner, J.",
  title         = "Deciphering the extreme X-ray variability of the nuclear transient eRASSt J045650.3−203750. A likely repeating partial tidal disruption event",
  year          = 2023,
  month         = 1,
  journal       = "A{\&}A",
  volume        = 669,
  pages         = "A75",
  doi           = "10.1051/0004-6361/202244805",
  archivePrefix = "arXiv",
  eprint        = "2208.12452",
  bibcode       = "2023A&A...669A..75L"
}

@article{2023ApJ...942L..33W,
  author        = "Wevers, T. and Coughlin, E.~R. and Pasham, D.~R. and Guolo, M. and Sun, Y. and Wen, S. and Jonker, P.~G. and Zabludoff, A. and Malyali, A. and Arcodia, R. and Liu, Z. and Merloni, A. and Rau, A. and Grotova, I. and Short, P. and Cao, Z.",
  title         = "Live to Die Another Day: The Rebrightening of AT 2018fyk as a Repeating Partial Tidal Disruption Event",
  year          = 2023,
  month         = 1,
  journal       = "ApJL",
  volume        = 942,
  pages         = "L33",
  doi           = "10.3847/2041-8213/ac9f36",
  archivePrefix = "arXiv",
  eprint        = "2209.07538",
  bibcode       = "2023ApJ...942L..33W"
}

@article{2023ApJ...957...12R,
  author        = "Ryu, Taeho and Krolik, Julian and Piran, Tsvi and Noble, Scott C. and Avara, Mark",
  title         = "Shocks Power Tidal Disruption Events",
  year          = 2023,
  month         = 11,
  journal       = "ApJ",
  volume        = 957,
  pages         = 12,
  doi           = "10.3847/1538-4357/acf5de",
  archivePrefix = "arXiv",
  eprint        = "2305.05333",
  bibcode       = "2023ApJ...957...12R"
}

@article{2023MNRAS.520.3549M,
  author        = "Malyali, A. and Liu, Z. and Rau, A. and Grotova, I. and Merloni, A. and Goodwin, A.~J. and Anderson, G.~E. and Miller-Jones, J.~C.~A. and Kawka, A. and Arcodia, R. and Buchner, J. and Nandra, K. and Homan, D. and Krumpe, M.",
  title         = "The rebrightening of a ROSAT-selected tidal disruption event: repeated weak partial disruption flares from a quiescent galaxy?",
  year          = 2023,
  month         = 4,
  journal       = "MNRAS",
  volume        = 520,
  pages         = "3549--3559",
  doi           = "10.1093/mnras/stad022",
  archivePrefix = "arXiv",
  eprint        = "2301.05501",
  bibcode       = "2023MNRAS.520.3549M"
}

@article{2024MNRAS.530.1866S,
  author        = "Sandoval, Astor and Riquelme, Mario and Spitkovsky, Anatoly and Bacchini, Fabio",
  title         = "Particle-in-cell simulations of the magnetorotational instability in stratified shearing boxes",
  year          = 2024,
  month         = 5,
  journal       = "MNRAS",
  volume        = 530,
  pages         = "1866--1884",
  doi           = "10.1093/mnras/stae959",
  archivePrefix = "arXiv",
  eprint        = "2308.12348",
  bibcode       = "2024MNRAS.530.1866S"
}

@article{2024MNRAS.532.1522J,
  author        = "Jacquemin-Ide, Jonatan and Rincon, Fran{\c c}ois and Tchekhovskoy, Alexander and Liska, Matthew",
  title         = "Magnetorotational dynamo can generate large-scale vertical magnetic fields in 3D GRMHD simulations of accreting black holes",
  year          = 2024,
  month         = 8,
  journal       = "MNRAS",
  volume        = 532,
  pages         = "1522--1545",
  doi           = "10.1093/mnras/stae1538",
  archivePrefix = "arXiv",
  eprint        = "2311.00034",
  bibcode       = "2024MNRAS.532.1522J"
}

@article{2024Natur.625..463S,
  author        = "Steinberg, Elad and Stone, Nicholas C.",
  title         = "Stream-disk shocks as the origins of peak light in tidal disruption events",
  year          = 2024,
  month         = 1,
  journal       = "Natur",
  volume        = 625,
  pages         = "463--467",
  doi           = "10.1038/s41586-023-06875-y",
  archivePrefix = "arXiv",
  eprint        = "2206.10641",
  bibcode       = "2024Natur.625..463S"
}
\ifboolexpr{bool{arxiv} or bool{local}}{\bibhang1.25em\printbibliography}{}

\end{document}